\documentclass[journal]{IEEEtran}

\ifCLASSINFOpdf

\else

\fi

\usepackage[utf8]{inputenc}
\usepackage{amssymb}
\usepackage{stfloats}
\usepackage{placeins}

\usepackage{overpic}
\usepackage[font=small]{caption}
\usepackage{amsmath}
\usepackage{comment}
\usepackage{mathtools}
\usepackage{booktabs}
\DeclarePairedDelimiter{\norm}{\lVert}{\rVert}
\usepackage{cite}
\usepackage{subfig}
\usepackage{lipsum}
\usepackage{graphicx}
\usepackage{textcomp,gensymb}

\usepackage{multicol}
\usepackage{algorithm}
\usepackage[noend]{algpseudocode}
\usepackage{tikz}
\usepackage[english]{babel}

\usepackage{amsthm}
\theoremstyle{plain}
\usepackage{dirtytalk}

\graphicspath{ {./images/} }

\usepackage{amsmath}

\usepackage{mathtools}

\usepackage{tabularx}
\newcolumntype{L}[1]{>{\raggedright\arraybackslash}p{#1}}
\newcolumntype{C}[1]{>{\centering\arraybackslash}p{#1}}
\newcolumntype{R}[1]{>{\raggedleft\arraybackslash}p{#1}}
\usepackage{multirow}
\usepackage{mathtools, cuted}
\newtheorem*{remark}{Remark}

\begin{document}

\title{Mitigating Intra-Cell Pilot Contamination in Massive MIMO: A Rate Splitting Approach}

\author{\IEEEauthorblockN{Anup~Mishra, \IEEEmembership{Graduate Student Member, IEEE}, Yijie~Mao, \IEEEmembership{Member, IEEE}, Christo Kurisummoottil Thomas, \IEEEmembership{Member, IEEE}, Luca~Sanguinetti, \IEEEmembership{Senior Member, IEEE} and Bruno~Clerckx, \IEEEmembership{Fellow, IEEE}\vspace{-0.5cm}}
\thanks{Manuscript received June 15, 2022; revised September 5, 2022; accepted October 23, 2022. The
associate editor coordinating the review of this article and approving it for publication was C.-K. Wen.}
\thanks{The authors Anup Mishra and Bruno Clerckx are with the Department of Electrical and Electronic Engineering, Imperial College London, London SW7 2AZ,
U.K. (e-mail: anup.mishra17@imperial.ac.uk; b.clerckx@imperial.ac.uk).}
\thanks{Yijie Mao is with the School of Information Science and Technology, ShanghaiTech University, Shanghai 201210, China (e-mail: maoyj@shanghaitech.edu.cn).}
\thanks{Christo Kurisummoottil Thomas is with the Bradley Department of Electrical and Computer Engineering,
Virginia Tech, Arlington, VA, USA, 22203 (email:christokt@vt.edu).}
\thanks{Luca Sanguinetti is with the Dipartimento di Ingegneria dell’Informazione,
University of Pisa, 65122 Pisa, Italy (e-mail: luca.sanguinetti@unipi.it).}}

\maketitle
\begin{abstract}
  Massive multiple-input multiple-output (MaMIMO) has become an integral part of the {fifth-generation} (5G) standard, and is envisioned to be further developed in beyond 5G {(B5G)} networks. With a massive number of antennas at the base station (BS), MaMIMO is best equipped to cater prominent use cases of B5G networks such as enhanced mobile broadband (eMBB), ultra-reliable low-latency communications (URLLC) and massive machine-type communications (mMTC) or combinations thereof. However, one of the critical challenges to this pursuit is the sporadic access behaviour of a massive number of devices in practical networks that inevitably leads to the conspicuous pilot contamination problem. Conventional linearly precoded physical layer strategies employed for downlink transmission in time division duplex (TDD) MaMIMO would incur a noticeable spectral efficiency (SE)  loss in the presence of this pilot contamination. In this paper, we aim to integrate a robust multiple access and interference management strategy named rate-splitting multiple access (RSMA) with TDD MaMIMO for downlink transmission and investigate its SE performance. We propose a novel downlink transmission framework of RSMA in TDD MaMIMO, devise a precoder design strategy and power allocation schemes to maximize different network utility functions. Numerical results reveal that RSMA is significantly more robust to pilot contamination and always achieves a SE performance that is equal to or better than the conventional linearly precoded MaMIMO transmission strategy.
\end{abstract}


\IEEEpeerreviewmaketitle

\section{Introduction}\label{Intro}

\IEEEPARstart{M}{assive} multiple-input multiple-output (MaMIMO) has been widely regarded as one of the key technologies in {fifth generation} ($5$G) communication\cite{Luca@MaMIMO2,emil@MaMIMO2}. With a large array of service antennas at the base station (BS), MaMIMO is capable of enhancing the spectral efficiency (SE), energy efficiency (EE) and robustness of multi-user multiple-input multiple-output (MIMO) networks \cite{emil@Luca,Marzetta@IntroMassiveMIMO,larson@marzetta2014}. A MaMIMO network can operate in both time-division duplex (TDD) and frequency-division duplex (FDD) modes. Due to the massive number of antennas, downlink (DL) channel state information (CSI) acquisition in the FDD mode incurs a huge training overhead, thereby decreasing the SE of the network\cite{emil@marzetta2016}. In contrast, by exploiting the reciprocity of the uplink (UL) and DL physical propagation channels, CSI acquisition is much simpler in the TDD mode. At the BS, the CSI is acquired through UL training and then utilized for DL transmission \cite{massivemimobook}. As a result, the training length is proportional to the number of user equipments (UEs) rather than the number of BS antennas, which significantly reduces the CSI overhead. This makes TDD the preferred mode of operation in MaMIMO networks \cite{emil@marzetta2016}. Even with a low CSI overhead, a TDD MaMIMO network is not without its limitations. In the UL, the CSI acquisition is preferably done by assigning orthogonal pilots to different UEs. However, due to the scarcity of orthogonal sequences, UEs are typically forced to use the same pilot for UL training, leading to the issue of pilot contamination \cite{Joe@MarzettaPilot,massivemimobook}. With the advent of $5$G, and expanded use cases in dense crowded scenarios and massive machine type communications (mMTC), the problem of pilot contamination gets further exacerbated. Within a cell, a massive number of UEs and their sporadic access behaviour make orthogonal scheduling of all UEs or allocation of orthogonal pilots to all UEs infeasible for transmission \cite{Emil@pracMaMIMO,carvalho@randomaccess,Elis@Emil,mishra2022ratesplitting}. For such scenarios, random access techniques are typically employed to serve active UEs in the network, where active UEs randomly select a pilot sequence from a small pool of orthogonal sequences for UL training. As a result, it is highly likely that multiple UEs may share the same pilot for UL training resulting in even severe \textit{intra-cell} pilot contamination\cite{carvalho@randomaccess,mishra2022ratesplitting}. 
\par The problem of pilot contamination is a major challenge in TDD MaMIMO and could lead to a significant SE loss. In fact, with uncorrelated Rayleigh fading channels, pilot contamination is known to be performance limiting \cite{massivemimobook}. To address the challenge, pilot contamination and its mitigation has been studied by a wide body of existing literature \cite{PC@channelest_nuemann,PC@channelest_Thang,kumar@ChestPC,Pilotreuse@saxena,Upadhya@superimposed,Mahyiddin@PC}. {To that end, \cite{Luca@MaMIMO2,emil@MaMIMO2} have proved that with spatially correlated channels and minimum mean-square error (MMSE) processing, the capacity of a MaMIMO network is asymptotically (with respect to the number of transmit antennas) unlimited despite pilot contamination}. Nevertheless, the SE of a TDD MaMIMO network is practically limited and with a finite number of antennas, incurs a considerable performance loss in the presence of pilot contamination\cite{Luca@MaMIMO2}. The undesired consequence of pilot contamination, {i.e.}, low quality statistically dependent channel estimates, is well documented in the literature\cite{massivemimobook}. Designing precoders based on such low quality CSI will result in severe multi-user interference in the DL and the network may become interference limited. 
\par One possible solution to deal with the DL multi-user interference that stems from the intra-cell pilot contamination would be to adopt the robust interference management strategy introduced in \cite{mao2017rate} named rate-splitting multiple access (RSMA). RSMA has emerged as a robust physical (PHY)-layer transmission strategy and is considered as a promising paradigm for multiple access in {beyond $5$G (B$5$G)} and {sixth-generation ($6$G)} networks\cite{mishra2022rate_tutorial,Onur@6G,mao2022ratesplitting}. RSMA has been realized in different forms for different multi-antenna settings\cite{mao2022ratesplitting}. The simplest form of RSMA is based on $1$-layer rate-splitting (RS)\footnote{Henceforth, 1-layer RS will be referred to as ‘RS’ for brevity.} which only requires one layer of successive interference cancellation (SIC) at each UE \cite{RSintro16bruno}. At the BS, RS splits the messages of UEs into two parts, a common part and a private part. The common parts of UEs' messages are combined together and encoded into common streams. The common streams are meant to be decoded by all UEs but not necessarily intended to all of them. The private parts are encoded independently into private streams and are meant to be decoded by the intended UE only (and treated as noise at the non-intended UEs). By adjusting the message split and power allocated to the common and private streams, RSMA allows to partially decode the interference and partially treat the interference as noise. From an information theoretic perspective, RSMA has been studied in-depth  \cite{mishra2021ratesplitting,chenxi2017bruno,jafar@davoodiMIMODoF,RSintro16bruno,RS2016hamdi,enrico2017bruno} and was shown to achieve the optimal degree of freedom (DoF) in multiple-input single-output (MISO) and MIMO broadcast channels (BC) with imperfect CSI \cite{jafar@davoodiMIMODoF,RS2016hamdi}. Motivated by the DoF optimality, communication theoretic performance of RSMA was investigated in \cite{mishra2021ratesplitting,mao2017rate,mao2018EE,hamdi2016robust,RS2016hamdi,mao2019beyondDPC}, and was shown to outperform the conventional multi-user MIMO strategy with linear precoders (also known as linearly precoded space division multiple access--SDMA) and power-domain non-orthogonal multiple access (PD-NOMA) in terms of SE and EE with imperfect CSI at the transmitter (CSIT) \cite{mishra2021ratesplitting,mao2017rate,mao2018EE}. Interestingly, RSMA has been investigated to deal with the deleterious effects of {mobility in MaMIMO\cite{onur@mobility}}, multi-user interference in FDD MaMIMO \cite{Minbo2016MassiveMIMO} and TDD cell-free MaMIMO \cite{mishra2022ratesplitting}, and to mitigate residual transceiver hardware impairments in TDD MaMIMO \cite{RS@TDDMaMIMO}. {To the best of our knowledge, employing RSMA to address the issue of pilot contamination in TDD MaMIMO has not been investigated yet.}
\par In this paper, motivated by the access behaviour of UEs in
$5$G, in turn, the need to address the challenge of pilot contamination and the merits of RSMA in the presence of imperfect CSI, we propose a general\footnote{The framework is general in the sense that it is valid for any channel estimation and precoder design technique. {The use of generalized RS from \cite{mao2017rate} instead of 1-layer RS is left for future work.}} DL transmission framework of RSMA in a single-cell TDD MaMIMO and investigate its performance as a PHY-layer strategy. Our main objective is to answer a simple question: \textit{Can RSMA help mitigate the deleterious effects of pilot contamination in TDD MaMIMO?}
\begin{figure}
\centering
\includegraphics[width=8.9cm,height=6.5cm]{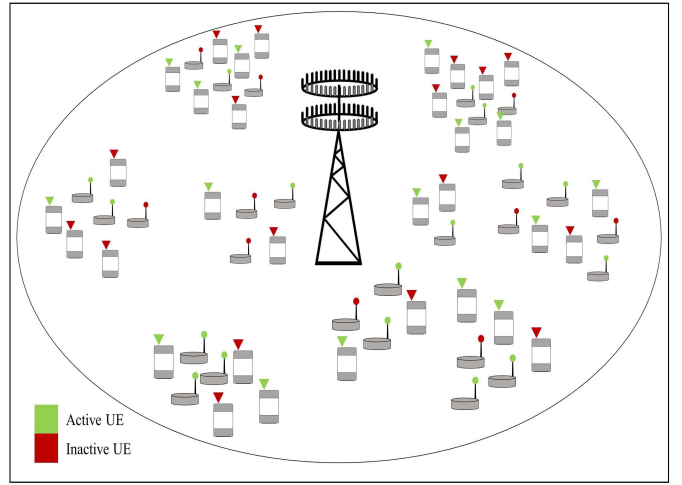}%
\caption{Illustration of massive access in a single-cell MaMIMO network for $5$G and beyond.}%
\label{fig:GenMassiveMIMO}
\end{figure}
\subsection{Contributions}
To investigate the efficacy of RSMA in TDD MaMIMO, we first analyze the DL performance of RSMA in TDD MaMIMO {with no pilot contamination}, i.e., all UEs using orthogonal pilot sequences for UL training. Next, to assess the performance in the  {presence of pilot contamination}, we assume that all UEs use the same pilot sequence for UL training.\footnote{The use of single pilot for UL transmission is not unprecedented in the literature of MaMIMO networks. \cite{single@pilot} considers the case where all UEs utilize a single pilot and investigates the UL performance of a conventional strategy in a distributed MaMIMO network.} The aim is to consider the best case (orthogonal pilots) and the worst case (same pilot) scenario in random access to determine the SE performance of RSMA and compare it with that of a conventional linearly precoded strategy. In this paper, we consider a single-cell MaMIMO network and therefore assume no inter-cell pilot contamination. {The main contributions of this paper are summarized as follows}:
\begin{itemize}
    \item We propose a novel system model employing RS in a TDD single-cell MaMIMO network. Based on the proposed system model, we derive the achievable SE expressions of RS and then obtain the capacity lower bound based on channel hardening (also known as hardening bound in \cite{massivemimobook}) for both the common and private streams. The derived hardening bounds are generalized for any UL channel estimation scheme and DL precoder design.
    \item To achieve a good SE performance with RS while keeping the computational complexity of precoder design low, we design the precoder for the common stream (common precoder) by maximizing the SE of the common stream, and precoders for the private streams by employing maximum ratio (MR) transmission. The design of the  common precoder solely depends on the channel statistics and therefore can be used for many coherence intervals.
    \item We propose three power allocation algorithms maximizing different network utilities for RS, namely, maximizing the sum-SE (MaxSum-SE), maximizing the product of signal to interference plus noise ratios (SINRs) (MaxSINR), and maximizing the minimum SE (MaxMin). For the MaxSum-SE problem, we propose a low-complexity heuristic algorithm for RS that aims to maximize the sum of SE of all UEs. For the MaxSINR problem, to maximize the product of SINRs of the common and private streams, we equivalently transform it into a geometric programming (GP) problem and solve it optimally. Finally, the MaxMin problem is formulated to maximize the worst-case SE among UEs. The MaxMin problem of RS is non-convex  and is solved using the proposed successive convex approximation (SCA)-based algorithm. 
    \item We compare the SE performance of  RS and NoRS\footnote{{Henceforth, a conventional MaMIMO transmission strategy implemented using multi-user linear precoding (also known as linearly precoded space division multiple access--SDMA) will be referred to as ‘NoRS’\cite{RS@TDDMaMIMO}.}} strategies for different pilot sharing scenarios, spatial correlation settings, and network topologies via extensive simulations. Numerical results illustrate the superiority of RS over NoRS in mitigating the harmful effects of pilot contamination, {i.e.,  degraded CSIT quality and statistically dependent channel estimates}, on DL transmission. {Since RSMA is more robust to imperfect CSIT and is a better interference management strategy in the DL, numerical results show that with pilot contamination it achieves better SE performance than NoRS in terms of both sum-SE and SE per UE}. This is the first work that proposes a general DL transmission framework of RSMA in TDD MaMIMO and investigates its efficacy as a pilot contamination mitigation strategy.
\end{itemize}
\subsection{Organization}
The rest of the paper is organized as follows. In Section \ref{Sysmod}, the system model is introduced. {In Section \ref{DL_RS}, SE expressions for the common and private streams are derived, and precoder design is proposed}. Section \ref{PowerAlloc} discusses the power allocation problems formulated for the three aforementioned network utility functions and describes their respective optimization methodologies in detail. Numerical results are illustrated and discussed in Section \ref{NumRes}, while Section \ref{Concl} concludes the paper. 

\subsection{Notations}
Matrices are denoted by boldface uppercase letters, column vectors are denoted by boldface lowercase letters and scalars are denoted by standard letters. Trace and determinant of matrix $\mathbf{A}$ are denoted by  $tr(\mathbf{A})$ and $\det(\mathbf{A})$, respectively. $diag(\mathbf{A})$ denotes the diagonal entries of the matrix. $\mathbf{A}^T$ and $\mathbf{A}^H$ denote the Transpose and Hermitian operators on matrix $\mathbf{A}$, respectively. Euclidean norm of  vector $\mathbf{a}$ is denoted as $\norm{\mathbf{a}}$. $\otimes$ denotes the Kronecker product and $vec(\mathbf{A})$ denotes vectorization of matrix $\mathbf{A}$. $\mathbb{E}_{X}\{Y\}$ is expectation of $Y$ with respect to random variable $X$. $\mathbb{C}^{M\times N}$ and $\mathbb{R}^{M\times N}$ denote the sets of all $M\times N$ dimensional matrices with complex-valued and real-valued entries, respectively. {A real Gaussian distribution with mean $\mu$ and variance $\sigma^{2}$ is denoted as $\mathcal{N}(\mu,\sigma^{2})$, whereas a circularly symmetric complex Gaussian (CSCG) distribution with mean $\mu$ and variance $\sigma^{2}$ is denoted as $\mathcal{CN}(\mu,\sigma^{2})$.}

\section{System Model}\label{Sysmod}
{We consider a single-cell MaMIMO network operating in the TDD mode with a BS equipped with $M$ transmit antennas simultaneously serving $K$ single-antenna  UEs in the same time-frequency resource block such that $M\gg1$ and $M/K >1$\cite{Luca@MaMIMO2}}. The UEs are indexed  by the set $\mathcal{K}=\{1,\ldots,K\}$. We use the standard block fading model, and in each block, the channel between UE $k$ and the BS, $\mathbf{g}_k\in\mathbb{C}^{M}$, is independently drawn from a block fading distribution as\cite{massivemimobook}
\begin{equation}\label{eq:channel_BSUE}
 \mathbf{g}_{k}=\sqrt{\beta_{k}}\mathbf{h}_{k} \sim \mathcal{CN}(\mathbf{0},\mathbf{R}_{k}), 
\end{equation}
where $\mathbf{R}_{k}\in \mathbb{C}^{M\times M}$ denotes the spatial correlation matrix. Gaussian distribution is used to model the small-scale fading variations $\mathbf{h}_{k}$, while $\mathbf{R}_{k}$ describes the large scale fading property which accounts for the path loss and shadowing effects. The normalized trace $\beta_{k}=\frac{1}{M}{tr}(\mathbf{R}_{k})$ denotes the average channel gain between the BS and UE $k$. Since UEs operate under the standard cellular MaMIMO TDD protocol, each coherence block consists of $\tau$ channel uses, whereof $\tau_p$ are used for UL pilot transmission, $\tau_u$ for UL data transmission and $\tau_d$ for DL data transmission such that $\tau = \tau_p + \tau_u + \tau_d$ \cite[Sec 2.1]{massivemimobook}. In this paper, we only consider UL pilot and DL data transmission and thus we set $\tau_u= 0$.
\par We assume that all UEs use a deterministic pilot sequence of length $\tau_p$. The pilot sequence of UE $k$ is denoted as $\{\boldsymbol{\phi}_{k}\in \mathbb{C}^{\tau_p}\mid \forall k\in \mathcal{K}\}$. We assume that each element in the pilot sequence has a magnitude $1/\sqrt{\tau_{p}}$ to obtain constant power levels and therefore $\norm{\boldsymbol{\phi}_{k}}^{2}=1,\,\forall k \in \mathcal{K}$. {We assume and denote $\rho_{\rm{ul}}$ as the average UL transmit power available at each UE}.\footnote{{The assumptions ensure that the channel estimation quality is not dependent on the pilot length $\tau_{p}$, and the difference in the channel estimation quality of two UEs solely depends on their respective $\beta$ values\cite{massivemimobook}. Such assumptions allow us to capture the impact of network topologies on pilot contamination, and in turn, analyze the SE performance of different transmission strategies with pilot contamination in different network settings.}} Following, the MMSE estimate of UE $k$ at the BS when a single pilot sequence is employed by all UEs is computed as \cite[Sec 3.2]{massivemimobook}
\begin{equation}\label{eq:channel_est}
\widehat{\mathbf{g}}_{k}= \mathbf{R}_{k}\mathbf{Q}^{-1}\big(\sum_{i=1}^{K}\mathbf{g}_{i} + \frac{1}{\sqrt{\rho_{\rm{ul}}}}\mathbf{n}_{t,k}\big)\sim \mathcal{CN}(\mathbf{0},\boldsymbol{\Phi}_{k}),
\end{equation}
where  $\mathbf{n}_{t,k}\sim \mathcal{CN}(\mathbf{0},\sigma_{\textrm{ul}}^{2}\mathbf{I}_{M})$, $\boldsymbol{\Phi}_{k}= \mathbf{R}_{k}\mathbf{Q}^{-1} \mathbf{R}_{k}$, and $\mathbf{Q}=\sum_{i \in \mathcal{K}}\mathbf{R}_{i}+\frac{\sigma_{\textrm{ul}}^{2}}{\rho_{\rm{ul}}}\mathbf{I}_M$. The channel estimate $\widehat{\mathbf{g}}_{k}$ and the channel estimation error $\widetilde{\mathbf{g}}_{k}=\mathbf{g}_{k}-\widehat{\mathbf{g}}_{k}$ are independent random variables with distributions $\mathcal{CN}(\mathbf{0},\boldsymbol{\Phi}_{k})$ and $\mathcal{CN}(\mathbf{0},\boldsymbol{\mathbf{R}}_{k}-\boldsymbol{\Phi}_{k})$, respectively. Since all UEs use the same pilot sequence, they contaminate each others' channel estimates, which makes their channel estimates statistically dependent. {Assuming $\mathbf{R}_{i}$ to be invertible, it follows that the channel estimate of UE $k$ at the BS can be written as \cite[eq (3.16)]{massivemimobook}},
\begin{equation}\label{eq:Dependent_Ch}
\widehat{\mathbf{g}}_{k}=\mathbf{R}_{k}\mathbf{R}_{i}^{-1}\widehat{\mathbf{g}}_{i},\;\; i \in \mathcal{K},
\end{equation}
and therefore, $\mathbb{E}\{\widehat{\mathbf{g}}_{i}\widehat{\mathbf{g}}_{k}^{H}\}=\mathbf{R}_{i}\mathbf{Q}^{-1}\mathbf{R}_{k}$ \cite{christo@ICC,massivemimobook}.
\par Similarly, if UEs are assigned orthogonal pilots for UL training, the channel estimate of UE $k$ is computed as
\begin{equation}\label{eq:channel_estortho}
\widehat{\mathbf{g}}_{k}= \mathbf{R}_{k}\mathbf{Q}_{k}^{-1}\big(\mathbf{g}_{k} + \frac{1}{\sqrt{\rho_{\rm{ul}}}}\mathbf{n}_{t,k}\big)\sim \mathcal{CN}(\mathbf{0},\boldsymbol{\Phi}_{k}),
\end{equation}
where $\boldsymbol{\Phi}_{k}=\mathbf{R}_{k}\mathbf{Q}_{k}^{-1} \mathbf{R}_{k}$ and  $\mathbf{Q}_{k}^{-1}=\mathbf{R}_{k}+\frac{\sigma_{\textrm{ul}}^{2}}{\rho_{\rm{ul}}}\mathbf{I}_M$. Since the pilots are orthogonal to each other, channel estimate of one UE is not contaminated by the channel of other UE, and thus the estimates are not statistically dependent.
\section{Rate-Splitting in DL Transmission}\label{DL_RS}
\begin{figure}
\centering
\includegraphics[width=8.9cm,height=5.4 cm]{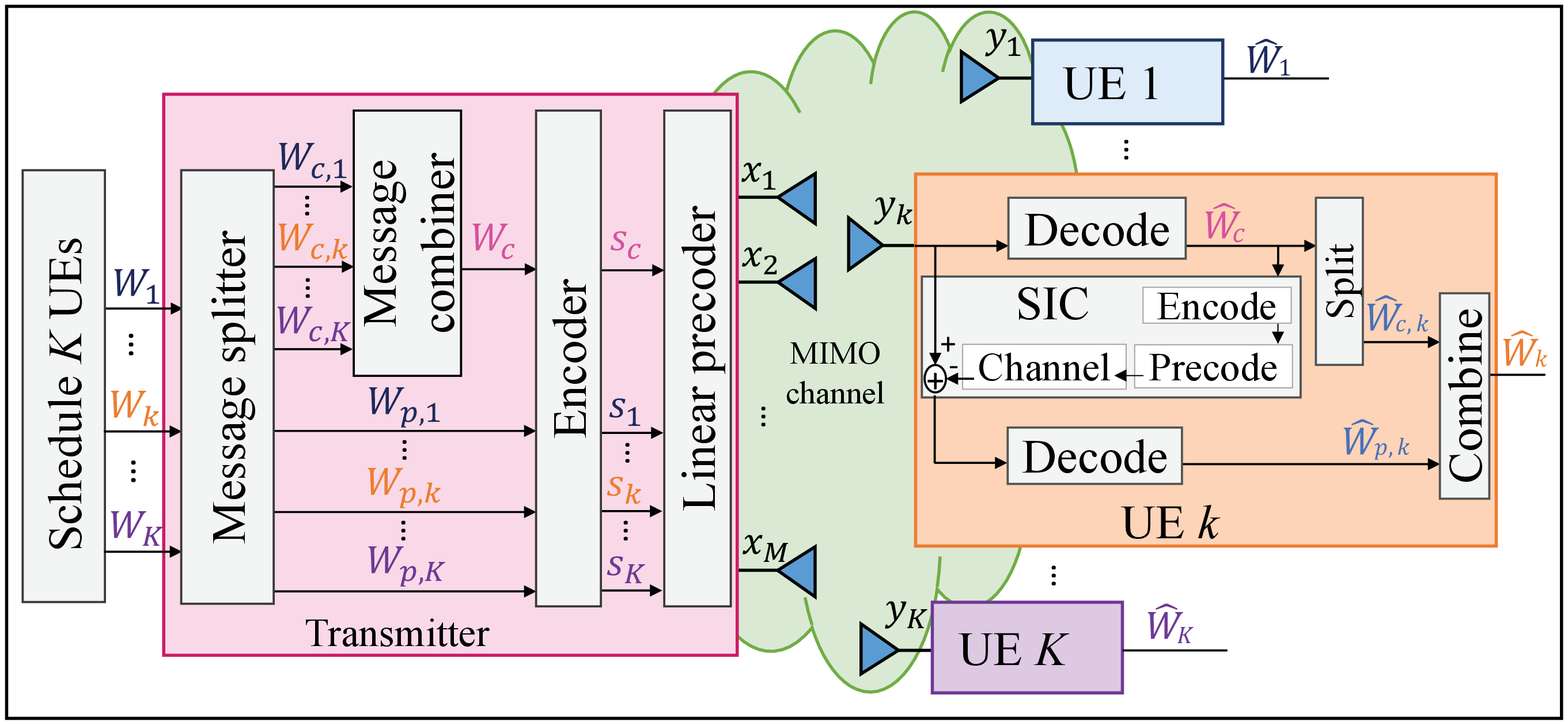}%
\caption{{K--UE DL transmission framework of RS\cite{mao2017rate}.}}%
\label{fig:GenFrameRS}
\end{figure}

For DL transmission, we use the RS strategy described in \cite{RSintro16bruno,mao2017rate} where the message of UE $k$, $\forall \,k \in\mathcal{K}$ denoted by ${W}_{k}$ is split into two parts, a common part ${W}_{c,k}$ and a private part ${W}_{p,k}$. The common parts of all UEs,  $\{{W}_{c,1},\ldots,{W}_{c,K}\}$, are combined together to form a single common message denoted as $W_{c}$ and then encoded into a single common stream ${s}_c\in \mathbb{C}$ using a common codebook such that $\mathbb{E}\{|{s}_{c}|^{2}\}=1$. The common stream is meant to be decoded by all UEs (but not necessarily intended to all of them). The private part of the message of UE $k$ is encoded independently into the private stream ${s}_{k}\in \mathbb{C}$ such that $\mathbb{E}\{|s_{k}|^{2}\}=1,\;\forall k \in \mathcal{K}$, and is meant to be decoded by the corresponding UE only. The DL transmission framework of the RS strategy with $K$ UEs is illustrated in Fig.~\ref{fig:GenFrameRS}. The resulting transmitted signal is written as
\begin{equation}\label{eq:trans_sig}
\mathbf{x}=\sqrt{\rho_c}\mathbf{w}_{c}{s}_{c} +\sum_{k=1}^{K}\sqrt{\rho_k}\mathbf{w}_{k}{s}_{k},
\end{equation}
where $\mathbf{w}_{c} \in \mathbb{C}^{M}$ is the precoder of the common stream such that $\mathbb{E}\{\norm{\mathbf{w}_{c}}^2\}=1$. $\mathbf{w}_{k} \in \mathbb{C}^{M}$ is the precoder for the private stream of UE $k$ such that $\mathbb{E}\{\norm{\mathbf{w}_{k}}^2\}=1, \forall k \in \mathcal{K}$ and it determines the spatial directivity of UE $k$. Note that the precoder normalization for both common and private precoders is taken as such for analytical tractability \cite[Sec 4.3]{massivemimobook}. The powers allocated to the common and private streams are denoted by $\rho_{c}$ and $\rho_{k},\,\forall k \in \mathcal{K}$, respectively. We define the DL transmit power constraint as 
\begin{equation}\label{eq:pow_cons}
\rho_{c} + \sum_{k=1}^{K} \rho_{k} \leq \rho_{\textrm{dL}},    
\end{equation}
where $\rho_{\textrm{dL}}$ is the total transmit power available at the BS for DL transmission. At the UE side, the received signal $y_{k}\in \mathbb{C}$ at UE $k$ is given by
\begin{equation}\label{eq:rx_sig}
y_{k} = \sqrt{\rho_c}\,\mathbf{g}_{k}^{H}\mathbf{w}_{c}{s}_{c} +\sum_{i=1}^{K}\sqrt{\rho_i}\,\mathbf{g}_{k}^{H}\mathbf{w}_{i}{s}_{i}+n_{k},
\end{equation}
where $n_{k}\in \mathcal{CN}(0,\sigma_{n,k}^{2})$ is the noise at UE $k$. Without loss of generality, we assume noise variances across UEs to be $\sigma_{n,k}^{2}=\sigma_{n}^{2},\, \forall k \in \mathcal{K}$. At UE $k$, first the common stream is decoded into $\widehat{W}_{c}$ by treating the interference from all private streams as noise. After decoding and successfully removing the common stream using SIC, UE $k$ decodes its own private stream into $\widehat{W}_{p,k}$ by treating the private streams of other UEs as noise. UE $k$ reconstructs its message by extracting $\widehat{W}_{c,k}$ from $\widehat{W}_{c}$, and combining it with $\widehat{W}_{p,k}$ to form $\widehat{W}_{k}$.
\subsection{Spectral Efficiency}
Assuming channel reciprocity within a coherence block, the BS then uses the estimates of UL channels to compute the precoders for DL data transmission. Since the UE is unaware\footnote{In TDD MaMIMO, due to asymptotic channel hardening, instantaneous CSI is not needed at the receiver and knowledge of statistical properties can be used to find good estimate of the channel \cite{NoPilot@Larsson} (and references there in). This assumption significantly reduces the resource (power and training duration) overhead and therefore is a widely used assumption in the literature of TDD MaMIMO\cite{massivemimobook,NoPilot@Larsson}.} of its exact channel, we assume that UE $k$ has knowledge of the ergodic effective precoded channels $\mathbb{E}\{\mathbf{g}_{k}^{H}\mathbf{w}_{c}\}$ and $\mathbb{E}\{\mathbf{g}_{k}^{H}\mathbf{w}_{k}\}$ \cite{massivemimobook}. { Under this assumption, only the part of the signal received over the ergodic effective precoded channel is treated as the true desired signal, and therefore the received signal at UE $k$ can be expressed as
\begin{equation}\label{eq:common_rx_E}
\begin{split}
y_{c,k}=&\sqrt{\rho_c}\,\mathbb{E}\{\mathbf{g}_{k}^{H}\mathbf{w}_{c}\}s_{c} + \sqrt{\rho_c}\,(\mathbf{g}_{k}^{H}\mathbf{w}_{c}-\mathbb{E}\{\mathbf{g}_{k}^{H}\mathbf{w}_{c}\}){s}_{c}\\+&\sum_{i=1}^{K}\sqrt{\rho_i}\,\mathbf{g}_{k}^{H}\mathbf{w}_{i}{s}_{i}+n_{k},
\end{split}
\end{equation}
where the first term in (\ref{eq:common_rx_E}) is the desired signal (for the common stream) over the known ergodic precoded channel $\mathbb{E}\{\mathbf{g}_{k}^{H}\mathbf{w}_{c}\}$, the second term is the desired signal over the unknown channel and the remaining terms are the interference from private streams of all UEs plus noise\cite{massivemimobook}}. After SIC\footnote{{Since the UE does not have perfect knowledge of the CSI, SIC of the common stream is not perfect. As a result, residual interference from the common stream still remains after SIC and is reflected in equation (\ref{eq:private_rx_E})}.} of the common stream, the received signal at UE $k$ becomes
\begin{equation}\label{eq:private_rx_E}
\begin{split}
y_{p,k}&=\sqrt{\rho_k}\,\mathbb{E}\{\mathbf{g}_{k}^{H}\mathbf{w}_{k}\}s_{k} + \sqrt{\rho_k}\,(\mathbf{g}_{k}^{H}\mathbf{w}_{k}-\mathbb{E}\{\mathbf{g}_{k}^{H}\mathbf{w}_{k}\}){s}_{k}\\
&+\sqrt{\rho_c}\,(\mathbf{g}_{k}^{H}\mathbf{w}_{c}-\mathbb{E}\{\mathbf{g}_{k}^{H}\mathbf{w}_{c}\}){s}_{c} + \sum_{i\neq k}^{K}\sqrt{\rho_i}\,\mathbf{g}_{k}^{H}\mathbf{w}_{i}{s}_{i}+n_{k}.
\end{split}
\end{equation}
The limited knowledge of the channel at UE makes it hard to characterize the DL SE for both the common and private stream. Therefore, we compute the lower bound of the ergodic capacity, known as the hardening bound, and obtain 
\begin{equation}\label{eq:SE_commonk}
\textrm{SE}_{c,k}=\frac{\tau_{d}}{\tau}\log(1+\gamma_{c,k}),
\end{equation}
\begin{equation}\label{eq:SE_private}
\textrm{SE}_{p,k}=\frac{\tau_{d}}{\tau}\log(1+\gamma_{p,k}),
\end{equation}
where $\gamma_{c,k}$ and $\gamma_{p,k}$ are the effective DL SINR lower bounds\footnote{Using Corollary 1.3, Theorem 4.6 (and its proof in Appendix C.3.6) of  \cite{massivemimobook} derives the lower bound of the DL ergodic channel capacity of a UE for the NoRS transmission strategy. The proof can be directly extended for RS by utilizing the effective SINR expressions (\ref{eq:SINR_C}) and (\ref{eq:SINR_P}) of common and private streams derived here and equation (1.9) of \cite{massivemimobook}.} of the common and private streams at UE $k$, respectively. With the expectations computed over channel realizations, $\gamma_{c,k}$ and $\gamma_{p,k}$ are given by
(\ref{eq:SINR_C}) and (\ref{eq:SINR_P}), respectively. The expectations of the effective precoded channels are with respect to the channel realizations and can be calculated using Monte Carlo simulations. As the common stream is decoded by all UEs, the achievable SE for the common stream (common SE) is defined as
\begin{equation}\label{eq:min_common}
\textrm{SE}_{c}=\frac{\tau_{d}}{\tau}\log(1+\gamma_{c}),
\end{equation}
where $\gamma_{c}=\min_{k\in\mathcal{K}}\gamma_{c,k}$. Note that the hardening bounds hold for any choice of channel estimator and precoder design. {Moreover, since the common message has common parts of the UEs' messages, we have {$\textrm{SE}_{{c}}=\sum_{k=1}^{K}C_{k}$}, where $C_{k}$ is the share of the common SE intended for UE $k$. Therefore, the total SE of UE $k$ is calculated as\cite{mao2017rate}
\begin{equation}\label{eq:Total_SE_K}
    \textrm{SE}_{k}=\textrm{SE}_{{p},k}+C_{k}, \forall k\in\mathcal{K}.
\end{equation}}
\begin{figure*}
\begin{equation}\label{eq:SINR_C}
\gamma_{c,k}=\frac{\rho_{c}|\mathbb{E}\{\mathbf{g}_{k}^{H}\mathbf{w}_c\}|^{2}}{\sum_{i=1}^{K}\rho_{i}\mathbb{E}\{|\mathbf{g}_{k}^{H}\mathbf{w}_i|^{2}\}+\rho_{c}(\mathbb{E}\{|\mathbf{g}_{k}^{H}\mathbf{w}_c|^{2}\}-|\mathbb{E}\{\mathbf{g}_{k}^{H}\mathbf{w}_c\}|^{2})+\sigma_{n}^{2}}, 
\end{equation}
\hrule
\end{figure*}
\begin{figure*}
\begin{equation}\label{eq:SINR_P}
\gamma_{p,k}=\frac{\rho_{k}|\mathbb{E}\{\mathbf{g}_{k}^{H}\mathbf{w}_k\}|^{2}}{\sum_{i=1}^{K}\rho_{i}\mathbb{E}\{|\mathbf{g}_{k}^{H}\mathbf{w}_i|^{2}\}-\rho_{k}|\mathbb{E}\{\mathbf{g}_{k}^{H}\mathbf{w}_k\}|^{2}+\rho_{c}(\mathbb{E}\{|\mathbf{g}_{k}^{H}\mathbf{w}_c|^{2}\}-|\mathbb{E}\{\mathbf{g}_{k}^{H}\mathbf{w}_c\}|^{2})+\sigma_{n}^{2}},
\end{equation}
\hrule
\end{figure*}
\vspace{-0.5cm}
\subsection{Precoder Design}\label{PrecoderDesign}
The  SE expressions in (\ref{eq:SE_private}) and (\ref{eq:min_common}) are general and can be utilized for any choice of common and private precoders. Due to the massive number of transmit antennas at the BS, high dimensional optimization of common and private precoders is infeasible in MaMIMO. Instead, low-complexity precoder design is desired for both common and private streams. Therefore, for private streams, we choose MR transmission precoders as they have low-complexity, achieve good SE performance with a high number of transmit antennas, and allow closed-form computation of the expectations in the SINR expressions\cite{christo@ICC,massivemimobook}. The MR precoder for the private stream of UE $k$ is defined as
\begin{equation}\label{eq:MR_Pre}
\mathbf{w}_{k}=\frac{\widehat{\mathbf{g}}_{k}}{\sqrt{\mathbb{E}\{\norm{\widehat{\mathbf{g}}_{k}}^2\}}}=\frac{\widehat{\mathbf{g}}_{k}}{\sqrt{tr(\boldsymbol{\Phi}_{k})}}.
\end{equation}
With MR precoders, the closed-form expectations $\mathbb{E}\{|\mathbf{g}_{k}^{H}\mathbf{w}_i|^{2}\}$ and $|\mathbb{E}\{\mathbf{g}_{k}^{H}\mathbf{w}_{k}\}|^{2}$, $\forall i,k\in\mathcal{K}$ are calculated as 
\begin{equation}\label{eq:Closedform_ExpSq_Pri}
|\mathbb{E}\{\mathbf{g}_{k}^{H}\mathbf{w}_k\}|^{2}=tr(\boldsymbol{\Phi}_{k}),
\end{equation}
\begin{equation}\label{eq:Closedform_SqExp_Pri}
\mathbb{E}\{|\mathbf{g}_{k}^{H}\mathbf{w}_{i}|^{2}\}=\frac{tr(\mathbf{R}_{k}\boldsymbol{\Phi}_{i})+|tr(\mathbf{R}_{k}\mathbf{Q}^{-1}\mathbf{R}_{i})|^2}{tr(\boldsymbol{\Phi}_{i})}.
\end{equation}
Note that only large scale fading coefficients are needed to calculate the closed-form expectations and, in turn, the hardening bounds.
\par Next we look at the design of the common precoder. Since the common stream is to be decoded by all UEs, considering equation (\ref{eq:min_common}), an ideal precoder for the common stream would be the one that maximizes the common SE. Consequently, the common precoder design problem can be formulated as 
\begin{subequations}\label{eq:common_precoder_opt}
\begin{align}
\max_{\mathbf{w}_{c}}&\;\;\min_{k}\;\;\;\;\gamma_{c,k},\\
&\;\; s.t.\;\;\;\; \mathbb{E}\{\norm{\mathbf{w}_{c}}^2\}=1.
\end{align}
\end{subequations}
Unfortunately, obtaining the optimal solution to problem (\ref{eq:common_precoder_opt}) is computationally demanding because of the beamforming gain uncertainty term in the denominator of the SINR expressions of the common stream, $\rho_{c}(\mathbb{E}\{|\mathbf{g}_{k}^{H}\mathbf{w}_c|^{2}\}-|\mathbb{E}\{\mathbf{g}_{k}^{H}\mathbf{w}_c\}|^{2})$, which makes problem (\ref{eq:common_precoder_opt}) intractable to solve. Moreover, such optimization will undesirably increase the time required for computing the common precoder. For ease of computation, a sub-optimal solution to problem (\ref{eq:common_precoder_opt}) can be obtained by assuming that the difference $(\mathbb{E}\{|\mathbf{g}_{k}^{H}\mathbf{w}_c|^{2}\}-|\mathbb{E}\{\mathbf{g}_{k}^{H}\mathbf{w}_c\}|^{2})$ is very small and can be neglected to make the problem tractable.\footnote{{The sole objective of assuming   beamforming gain uncertainty  $(\mathbb{E}\{|\mathbf{g}_{k}^{H}\mathbf{w}_c|^{2}\}-|\mathbb{E}\{\mathbf{g}_{k}^{H}\mathbf{w}_c\}|^{2})$ of the common stream at UE $k$ to be negligible is to make problem (\ref{eq:common_precoder_opt}) tractable to solve. Although such an assumption will result in sub-optimal common precoder design, the design complexity is much reduced. Note that the assumption of considering beamforming gain uncertainty negligible is only restricted to the common precoder design at the BS. The assumption is not a part of the system model, or SE calculations at the BS, or SE calculations at the UEs.}} Following this assumption, problem (\ref{eq:common_precoder_opt}) can be formulated as
\begin{equation}\label{eq:common_precoder_opt1}
\begin{split}
\max_{\mathbf{w}_{c}}&\;\;\min_{k}\;\;\;\; \pi_{k}\,|\mathbb{E}\{\mathbf{g}_{k}^{H}\mathbf{w}_{c}\}|^{2}\\
&\;\;s.t.\;\;\;\;\;\;\mathbb{E}\{\norm{\mathbf{w}_{c}}^2\}=1,
\end{split}
\end{equation}
where 
\begin{equation}\label{eq:pi_k}
\pi_k=\rho_{c}\left({\sum_{i=1}^{K}\rho_{i}\mathbb{E}\{|\mathbf{g}_{k}^{H}\mathbf{w}_i|^{2}\}+\sigma_{n}^{2}}\right)^{-1}.
\end{equation}
We consider a weighted MR approach and design the common precoder in the span of subspace of the estimated channel vectors $\{\widehat{\mathbf{g}_{i}} \mid \,\forall i\in\mathcal{K}\}$ at the BS as
\begin{equation}\label{eq:common_pre_hue}
\mathbf{w}_{c}=\Omega\sum_{i=1}^{K}a_{i}\widehat{\mathbf{g}}_{i},
\end{equation}
{where $a_{i}$ is the weight assigned to the estimate of UE $i,\,i\in\mathcal{K}$ and} $\Omega$ is the scaling factor required to satisfy the constraint $\mathbb{E}\{\norm{\mathbf{w}_{c}}^2\}=1$.\footnote{Reference {\cite{Minbo2016MassiveMIMO} also adopts the weighted MR approach for common precoder design of RSMA in FDD MaMIMO. In \cite{Minbo2016MassiveMIMO}, the closed-form computation of the coefficients $a_{i},\,\forall i\in\mathcal{K}$ is done based on the instantaneous SINR expressions. Moreover, DL training allows the UE to have the knowledge of the effective precoded channel. In contrast, we consider no DL training at the UE, and calculate common precoder coefficients utilizing the hardening bounds of the common and private SINRs.}} By substituting (\ref{eq:common_pre_hue}) into $\mathbb{E}\{\mathbf{g}_{k}^{H}\mathbf{w}_{c}\}$, we rewrite (\ref{eq:common_precoder_opt1}) as
\begin{subequations}\label{eq:common_precoder_opt2}
\begin{align}
\max_{\{a_{k}\mid \forall k\in \mathcal{K}\}}&\min_{k}\;\;\;\Omega^{2}\,\pi_k \,\left|\sum_{i=1}^{K}a_{i}\mathbf{U}(i,k)\right|^2,\\
&s.t.\;\;\;\mathbb{E}\{\norm{\mathbf{w}_{c}}^2\}=1,
\end{align}
\end{subequations}
where  
\begin{equation}\label{eq:matrix_U}
    \mathbf{U}(i,k)=\mathbb{E}\{\widehat{\mathbf{g}}_{k}^{H}\widehat{\mathbf{g}}_{i}\},\,\forall i,k\in\mathcal{K}.
\end{equation}
For simplicity, we ignore $\Omega^{2}\pi_{k}$ in (\ref{eq:common_precoder_opt2}). By introducing an auxiliary variable $t$, we obtain the convex form of the objective and constraints of problem (\ref{eq:common_precoder_opt2}) and equivalently transform it into the following convex optimization problem: 
\begin{subequations}\label{eq:commonpre_convexprob}
\begin{align}
\max_{\mathbf{a},\,t>0}&\;\;\;\;\;\;\;t,\\
&s.t. \;\;\; \mathbf{a}^{T}\mathbf{U}(:,k) \geq t,\;\;\; \forall k \in \mathcal{K},
\end{align}
\end{subequations}
where $\mathbf{a}=[a_{1},\ldots,a_{K}]^{T}$. {Problem (\ref{eq:commonpre_convexprob}) is equivalent to (\ref{eq:common_precoder_opt2}) as it optimizes the weights, $\mathbf{a}$, with the aim of maximizing the minimum achievable SE of the common stream at each UE, $t$, which corresponds to the objective of (\ref{eq:common_precoder_opt2})}. Consequently, we solve problem (\ref{eq:commonpre_convexprob}) to obtain $\mathbf{a}^{*}$ which is then used to compute the optimal common precoder $\mathbf{w}_{c}^{*}$. For the scenario of every UE using the same pilot for UL channel estimation, we have $\mathbf{U}(i,k)=tr(\mathbf{R}_{i}\mathbf{Q}^{-1}\mathbf{R}_{k})$ and the common precoder is computed as 
\begin{equation}\label{eq:Opt_commonPre}
\mathbf{w}_{c}^{*}=\frac{\sum_{i=1}^{K}a_{i}^{*}\widehat{\mathbf{g}}_{i}}{\sqrt{\sum_{i=1}^{K}\sum_{j=1}^{K}a_{i}^{*}a_{j}^{*}tr(\mathbf{R}_{i}\mathbf{Q}^{-1}\mathbf{R}_{j})}}.  
\end{equation}
Using (\ref{eq:Opt_commonPre}), we compute the closed-form expressions of the expectations $\mathbb{E}\{\mathbf{g}_{k}^{H}\mathbf{w}_{c}^{*}\}$ and $\mathbb{E}\{\mathbf{|g}_{k}^{H}\mathbf{w}_{c}^{*}|^{2}\}$ which are solely dependent on the channel statistics. Appendix A specifies the derivation of these expectations.
\par {With orthogonal pilots, we have  $\{\mathbf{U}(i,k)=tr(\boldsymbol{\Phi}_{i})\mid i=k\}$ and $\{\mathbf{U}(i,k)=0 \mid \forall k\neq i\}$. To avoid redundancy, we do not elaborate on the common precoder calculation for orthogonal pilots. Similarly, the common precoder design problem in (\ref{eq:common_precoder_opt1}) can be solved for any pilot sharing scenario by calculating the corresponding value of $\mathbf{U}$ using (\ref{eq:matrix_U}).} The computation of common precoder coefficients $\mathbf{a}$, solely depends on the large scale fading coefficients. Therefore, $\mathbf{a}$ only needs to be calculated once. Consequently, the common precoder can be calculated using $\mathbf{a}$ and (\ref{eq:common_pre_hue}) for many coherence intervals, until the channel statistics change. Such precoder design keeps the overall time and computational burden very low.

\section{Power Allocation}\label{PowerAlloc}
In this section, we aim to obtain the power allocation coefficients $\boldsymbol{\rho}=\{\rho_c, \rho_1,\ldots,\rho_K\}$ for the common and private streams of RS by considering different utility functions as optimization objectives.  We consider utility functions that capture the aggregate SE performance of the network, fairness in the SE performance of UEs, and strike a balance between aggregate SE performance and fairness. We define a network utility function as $U(\textrm{SE}_{1},\ldots,\textrm{SE}_{K})$ which takes SE of UEs as input, and returns a scalar that measures the utility as the output. To that end, following equations (\ref{eq:SE_private})-(\ref{eq:Total_SE_K}), the utility functions for RS can be written as
\begin{equation}\label{eq:Gen_Utility_Fns}
   U(\textrm{SE}_{1},\ldots,\textrm{SE}_{K}):\begin{dcases}
    \textrm{SE}_{c} + \sum_{k=1}^{K} \textrm{SE}_{p,k},&\textrm{Sum-SE}\\ 
    \left(\prod_{k=1}^{K}\gamma_{p,k}\right)\gamma_{c},&\textrm{Product\,of\,SINRs}\\
   \min_{k\in\mathcal{K}}\;\textrm{SE}_{p,k}+C_{k},&\textrm{Minimum\;SE},
  \end{dcases}
\end{equation}
where $\gamma_{c}$ and $\gamma_{p,k},\,\forall k\in \mathcal{K}$ are effective SINRs of the common and private streams SEs, respectively. We formulate and solve power allocation problems of RS to maximize the three different utility functions in (\ref{eq:Gen_Utility_Fns}), i.e., 1) maximizing sum-SE (MaxSum-SE), 2) maximizing product of SINRs (MaxSINR) and 3) maximizing the minimum SE (MaxMin). {All power allocation schemes designed in this paper will hold for any channel estimation and precoder design method. Moreover, each power allocation scheme is solely dependent on the channel statistics and therefore can be used for many coherence intervals}. We define the utility function maximization problem as
\begin{equation}\label{eq:Utility_max}
\begin{split}
\max_{\boldsymbol{\rho}}&\;\;\; U(\textrm{SE}_{1},\ldots,\textrm{SE}_{K}) ,\\
&s.t.\;\;\; \rho_c+\sum_{i=1}^{K}\rho_{i}\leq \rho_{\textrm{dL}}.
\end{split}
\end{equation}
 For simplicity, we first rewrite equations (\ref{eq:SINR_C}) and (\ref{eq:SINR_P}) as
\begin{equation}\label{eq:SINR_C1}
\gamma_{c,k}=\frac{\rho_{c}a_{c,k}}{\sum_{i=1}^{K}\rho_{i}b_{ki}^{c}+\rho_{c}I_{c,k}+\sigma_{n}^{2}}, 
\end{equation}
\begin{equation}\label{eq:SINR_P1}
\gamma_{p,k}=\frac{\rho_{k}a_{p,k}}{\sum_{i=1}^{K}\rho_{i}b_{ki}^{p}+\rho_{c}I_{c,k}+\sigma_{n}^{2}}, \end{equation}
respectively, where
\begin{equation}\label{eq:a_cpk}
\begin{split}
a_{c,k}&=|\mathbb{E}\{\mathbf{g}_{k}^{H}\mathbf{w}_c\}|^{2},\,\forall k\in\mathcal{K},\\
a_{p,k}&=|\mathbb{E}\{\mathbf{g}_{k}^{H}\mathbf{w}_k\}|^{2},\,\forall k\in\mathcal{K},\\    
\end{split}    
\end{equation}
\begin{equation}\label{eq:B_kipc}
\begin{split}
  b_{ki}^{c}&=\mathbb{E}\{|\mathbf{g}_{k}^{H}\mathbf{w}_i|^{2}\},\,\forall k\in\mathcal{K},\\
  b_{ki}^{p}&=
    \begin{cases}
      \mathbb{E}\{|\mathbf{g}_{k}^{H}\mathbf{w}_i|^{2}\}, &i\neq k,\,\forall\,i,k\in\mathcal{K},\\
      \mathbb{E}\{|\mathbf{g}_{k}^{H}\mathbf{w}_k|^{2}\}-|\mathbb{E}\{\mathbf{g}_{k}^{H}\mathbf{w}_k\}|^{2}, &i=k,\,\forall\, i,k\in\mathcal{K},
    \end{cases}  
\end{split}
\end{equation}
and
\begin{equation}\label{eq:ICK}
I_{c,k}=\mathbb{E}\{|\mathbf{g}_{k}^{H}\mathbf{w}_c|^{2}\}-|\mathbb{E}\{\mathbf{g}_{k}^{H}\mathbf{w}_c\}|^{2},\,\forall k\in\mathcal{K}.    
\end{equation}
Note that, while the channel estimates and the closed-form expressions of the expectations depend on whether UEs are using the same pilot or orthogonal pilots, the power allocation algorithms are not affected by the choice of pilot sequences or precoder design\footnote{While the power allocation algorithms themselves are unaffected by the choice of pilots or precoders, the power allocation between the common and private streams will be influenced by both.}. As aforementioned, only large scale fading characteristics are used to design the power allocation algorithms of RS in TDD MaMIMO. Therefore, similar to the NoRS strategy, the advantage of being able to design complex yet feasible power allocation schemes which can be used for multiple coherence blocks is retained for RS as well with our proposed transmission framework. In the following, we formulate the three power allocation problems of RS in TDD MaMIMO and specify the algorithms proposed to solve the corresponding problems.
\subsection{Maximizing sum-SE (MaxSum-SE)}
For any given channel estimation technique, precoding scheme and values of $a_{c,k}, a_{p,k}, b_{ki}^{c}, b_{ki}^{p}$ and $I_{c,k}$, the sum-SE with the RS transmission strategy can be written as
\begin{equation}\label{eq:SumSE_eq}
\textrm{SE}=\textrm{SE}_{c}+\sum_{k=1}^{K}\textrm{SE}_{p,k},
\end{equation}
where $\textrm{SE}_{c}$ and $\textrm{SE}_{p,k}$ are defined in (\ref{eq:min_common}) and (\ref{eq:SE_private}), respectively. The MaxSum-SE problem can therefore be formulated as 
\begin{subequations}\label{eq:MaxSumSE}
\begin{align}
\max_{\boldsymbol{\rho}}&\;\;\; \textrm{SE}_{c}(\boldsymbol{\rho})+\sum_{k=1}^{K}\textrm{SE}_{p,k}(\boldsymbol{\rho}),\\
&s.t.\;\;\; \rho_{c}+\sum_{i=1}^{K}\rho_{i}\leq \rho_{\textrm{dL}}.
\end{align}
\end{subequations}
We consider a heuristic low-complexity power allocation strategy to maximize the sum-SE where power allocated to the common stream is $\rho_{c}=(1-\zeta)\rho_{\textrm{dL}}$, and the power allocated to each private stream is $\rho_{k}={\zeta\rho_{\textrm{dL}}}/{K},\, \forall k\in\mathcal{K}$, where $\zeta \in [0,1]$ is a scalar. An exhaustive search for $\zeta$ is then considered to obtain the power allocation between the common and private streams for which the sum-SE is maximum. Such an approach does not require optimization techniques and maintains a low complexity, which is desirable to many potential use cases of future wireless networks. Algorithm \ref{alg:MaxSumSE_alg} outlines the MaxSum-SE power allocation algorithm of RS  that gives the power allocation between the common stream and private streams, and ultimately obtains the sum-SE. {To be thorough, the power allocation coefficients in (\ref{eq:MaxSumSE}) are also obtained using an optimization technique detailed in Appendix \ref{App_B}, and a comparison with Algorithm \ref{alg:MaxSumSE_alg} is provided in terms of sum-SE and complexity.} 
\begin{algorithm}
\caption{MaxSum-SE}\label{alg:MaxSumSE_alg}
\begin{algorithmic}[1]
\State $\mathbf{Initialize}\:\: n\leftarrow 0,\;\zeta \leftarrow 0,\;\Delta\leftarrow 0.05,\;\textrm{SE}^{[n]}$
\State $\mathbf{Iterate}$
\State $\;n\leftarrow n+1$;
\item[] $\;Obtain\;\;\;\rho_{c}\leftarrow (1-\zeta)\rho_{\textrm{dL}}\;\textrm{and}\; \;\rho_{k}\leftarrow \frac{\zeta\rho_{\textrm{dL}}}{K},\, \forall k\in\mathcal{K}$;
\item[] $\;Calculate\;\textrm{SE}^{[n]}\;\textrm{using} \;(\ref{eq:SINR_C1}),\;(\ref{eq:SINR_P1})\;\textrm{and}\;(\ref{eq:SumSE_eq})$;
\State $\mathbf{Update}\; \textrm{SE}^{[n]} \leftarrow \max(\textrm{SE}^{[n]},\textrm{SE}^{[n-1]}),\;\zeta \leftarrow \zeta+\Delta$ 
\State $\mathbf{Until}\;\;\;\zeta=1$
\end{algorithmic}
\end{algorithm}
\par {\textbf{Complexity}: From Algorithm \ref{alg:MaxSumSE_alg}, we can discern that the computational complexity of the algorithm depends on $\Delta$ and is of the order of $\mathcal{O}(\Delta^{-1})$}.
\subsection{Maximizing product of SINRs (MaxSINR)}
We next consider MaxSINR\footnote{Maximizing product of SINRs aims to maximize the sum-SE where ``${1+}$" term is neglected in every SE expression. While ignoring ``${1+}$" term has a minuscule affect on UEs with high SINRs, UEs with lower SINRs have their SEs underestimated\cite{massivemimobook}. {As a result, the objective of maximizing the product of SINRs will lead to higher SE for weaker UEs. Moreover, owing to  the objective function, each UE is guaranteed a non-zero SE thereby providing more fairness compared to the MaxSum-SE power allocation scheme.}} power allocation which aims to strike a balance between maximizing the sum-SE and maintaining fairness in the network. For any given channel estimation, precoding scheme and values of $a_{c,k}, a_{p,k}, b_{ki}^{c}, b_{ki}^{p}$ and $I_{c,k}$, MaxSINR utility function maximization problem of RS can be written as 
\begin{subequations}\label{eq:MaxSINR_1}
\begin{align}
\max_{\gamma_{c},\,\boldsymbol{\rho}}&\;\;\; \left(\prod_{i=1}^{K}\gamma_{p,i}\right)\gamma_{c}\\
s.t.&\;\;\;\gamma_{c}\leq \gamma_{c,k},\,\forall k\in\mathcal{K},\\
&\;\;\; \rho_{c} + \sum_{i=1}^{K} \rho_{i} \leq \rho_{\textrm{dL}},
\end{align}
\end{subequations}
where $\gamma_{c,k}$ and $\gamma_{p,k}$ are defined in (\ref{eq:SINR_C1}) and (\ref{eq:SINR_P1}), respectively. Constraint (\ref{eq:MaxSINR_1}b) ensures that the common stream is decodable at all UEs. Since problem (\ref{eq:MaxSINR_1}) is non-convex, we transform it into an equivalent GP form by introducing auxiliary variables 
$o_{c}$ and $\mathbf{o}_{p}=[o_{p,1},\ldots,o_{p,K}]$ to respectively represent the common SINR and the private SINR vector such that
\begin{subequations}\label{eq:SINRAux_cp}
\begin{align}
o_{p,k}\big(\sum_{i=1}^{K}\rho_{i}b_{ki}^{p}+\rho_{c}I_{c,k}+\sigma_{n}^{2}\big)&\leq \rho_{k}a_{p,k},\,\forall k\in\mathcal{K},\\ 
o_{c}\big(\sum_{i=1}^{K}\rho_{i}b_{ki}^{c}+\rho_{c}I_{c,k}+\sigma_{n}^{2}\big)&\leq \rho_{c}a_{c,k},\,\forall k\in\mathcal{K}.
\end{align}
\end{subequations}
Using (\ref{eq:SINRAux_cp}a) and (\ref{eq:SINRAux_cp}b), we transform problem (\ref{eq:MaxSINR_1}) equivalently into a GP problem given by
\begin{subequations}\label{eq:MaxSINR_2}
\begin{align}
\max_{\boldsymbol{\rho},\mathbf{o}_{p},o_{c}} &\;\;\;\;\;\;\;(\prod_{i=1}^{K}o_{p,i})o_{c}\\ 
&s.t.\;\;o_{p,k}\;\frac{\big(\sum_{i=1}^{K}\rho_{i}b_{ki}^{p}+\rho_{c}I_{c,k}+\sigma_{n}^{2}\big)}{\rho_{k}a_{p,k}}\leq 1,\; \forall k\in\mathcal{K},\\
&\;\;\;\;\;\;\;\;o_{c}\;\frac{\big(\sum_{i=1}^{K}\rho_{i}b_{ki}^{c}+\rho_{c}I_{c,k}+\sigma_{n}^{2}\big)}{\rho_{c}a_{c,k}}\leq 1,\; \forall k\in\mathcal{K},\\
&\;\;\;\;\;\;\;\;\rho_{c}+\sum_{i=1}^{K}\rho_{i}\leq {\rho}_{\textrm{dL}}.
\end{align}    
\end{subequations}
The objective function and the constraints in problem (\ref{eq:MaxSINR_2}) are posynomials making it a GP problem \cite{Boyd@2004}. We use CVX, a tool used to solve disciplined convex programs in Matlab for finding the solution of the above GP problem and obtaining the optimal power allocation $\boldsymbol{\rho}^{*}$ that maximizes the product of SINRs utility function for RS in TDD MaMIMO. Algorithm \ref{alg:MaxSINR_alg} outlines the MaxSINR power allocation algorithm of RS. It should be highlighted that in a GP, there is an implicit constraint that the optimization variables are positive, i.e., here $\rho_{i}>0,\,\forall i\in\mathcal{K}$ and $\rho_{c}>0$\cite{Boyd@2004}. This non zero power allocation to all the streams assures a non zero SE to each UE thereby ensuring fairness in the network. 
\begin{algorithm}
\caption{MaxSINR:\;GP Algorithm}\label{alg:MaxSINR_alg}
\begin{algorithmic}[1]
\State $\mathbf{Declare}\:\: o_{c},\;\mathbf{o}_{p}$
\item[]$\;\;{Transform}\; (\ref{eq:MaxSINR_1})\;\textrm{into}\;\textrm{GP}\;\textrm{problem}\;(\ref{eq:MaxSINR_2})$
\item[]$\;\;\;\;\;\;\;\;\;\;\;\;\;\;\;\;\;\;\;\;\;\;\textrm{using}\;(\ref{eq:SINRAux_cp}a)\;\textrm{and}\;(\ref{eq:SINRAux_cp}b)$;
\State $\mathbf{Solve}\;(\ref{eq:MaxSINR_2})\;\textrm{using}\;\textrm{CVX}\;\textrm{to}\;\textrm{obtain}\;\boldsymbol{\rho}$ 
\State $\mathbf{Calculate}\;\;\textrm{SE}_{c}\;\textrm{and}\;\textrm{SE}_{p,k}$
\end{algorithmic}
\end{algorithm}
\par {\textbf{Complexity}: Since problem (\ref{eq:MaxSINR_2}) has $2(K+1)$ variables and $2K+1$ constraints, based on \cite{Boyd@2004}, the computational complexity of solving problem (\ref{eq:MaxSINR_2}) is of the order of 
\begin{equation}\label{eq:maxSINR_complexity}
    \mathcal{O}\left(\max\left\{8\left(K+1\right)^{3},\,F_{1}\right\}\right),
\end{equation}
where $F_{1}$ is the cost of evaluating the first and second derivatives of the objective and constraint functions in (\ref{eq:MaxSINR_2}).}
\subsection{Maximizing the minimum SE (MaxMin)}
MaxSINR optimization, as aforementioned, aims to strike a balance between maximizing the sum-SE and fairness among UEs. We next consider MaxMin optimization problem whose objective is to achieve the maximum user fairness. The MaxMin problem of RS for any given channel estimation, precoding scheme and values of $a_{c,k}, a_{p,k}, b_{ki}^{c}, b_{ki}^{p}$ and $I_{c,k}$ can be expressed as
\begin{subequations}\label{eq:MaxMin_1}
\begin{align}
\max_{\boldsymbol{\rho},\mathbf{c}}&\;\min_{k}\;\;\;\; \textrm{SE}_{p,k}+C_{k},\\
&s.t.\;\;\;\;\;\;\;C_{1}+\ldots+C_{K}\leq \textrm{SE}_{c,k},\;\; \forall k \in \mathcal{K},\\
&\;\;\;\;\;\;\;\;\;\;\;\;\rho_{c} + \sum_{i=1}^{K} \rho_{i} \leq \rho_{\textrm{dL}},\\
&\;\;\;\;\;\;\;\;\;\;\;\; \mathbf{c}\geq \mathbf{0}, 
\end{align}
\end{subequations}
where $\mathbf{c}=[C_{1},\ldots,C_{K}]$ is the common SE vector with $C_{k}$ being the share of the common SE allocated to UE $k$ such that $\textrm{SE}_{c}=\sum_{i=1}^{K}C_{i}$. Constraint (\ref{eq:MaxMin_1}b) ensures that the common stream is decoded at all UEs. Our aim is to jointly optimize the power allocated to the common stream $\rho_c$, private streams $\rho_{i},\,\forall i\in\mathcal{K}$, and the common SE vector $\mathbf{c}$. 
\par The MaxMin problem (\ref{eq:MaxMin_1}) described above is a non-convex problem due to the presence of logarithmic and fractional SE expressions $\textrm{SE}_{p,k}$ and $\textrm{SE}_{c,k}$. Motivated by the SCA algorithm adopted in \cite{mao2018EE,Lina@SCA}\footnote{Both \cite{mao2018EE,Lina@SCA} formulate the SCA algorithm for precoder design in multi-user MISO networks with perfect CSIT.}, we propose a SCA-based power allocation algorithm to solve problem (\ref{eq:MaxMin_1}). To achieve the best possible performance, we introduce auxiliary variables to transform the MaxMin problem into its equivalent form and approximate the transformed problem into convex sub-problems, which are solved iteratively until convergence. Next, we delineate the transformation, approximations, and procedure to solve the transformed problem. 
\par We introduce an auxiliary variable $t$, vectors $\boldsymbol{\alpha_{c}}=[\alpha_{c,1},\ldots,\alpha_{c,K}]$, and $\boldsymbol{\alpha_{p}}=[\alpha_{p,1},\ldots,\alpha_{p,K}]$ representing the minimum SE, SE of the common stream at UEs and private SEs of UEs, respectively. Similarly, we introduce $\mathbf{r}_{c}=[r_{c,1},\ldots,r_{c,K}]$ and $\mathbf{r}_{p}=[r_{p,1},\ldots,r_{p,K}]$ representing $1$ plus SINR values of the common and private streams of UEs, respectively. As a result, problem (\ref{eq:MaxMin_1}) is equivalently transformed as
\begin{subequations}\label{eq:MaxMin_2}
\begin{align}
\max_{\substack{\boldsymbol{\rho},\mathbf{c},\boldsymbol{\alpha}_{p},\boldsymbol{\alpha}_{c},\\\mathbf{r}_{p},\mathbf{r}_{c},t}}&\;\;\; t\\ 
s.t.\;&\alpha_{p,k}+C_k \geq t,\;\;\; \forall k\in\mathcal{K},\\ 
&\alpha_{c,k} \geq \sum_{k\in\mathcal{K}}C_{k},\;\;\; \forall k\in\mathcal{K},\\ 
& r_{p,k} - 2^{\frac{\tau}{\tau_{d}}\alpha_{p,k}} \geq 0 \;\;\; \forall k\in\mathcal{K},\\ 
& r_{c,k} - 2^{\frac{\tau}{\tau_{d}}\alpha_{c,k}} \geq 0 \;\;\; \forall k\in\mathcal{K},\\ 
& \frac{\rho_c a_{c,
k}}{\sum_{i=1}^{K}\rho_i b_{ki}^{c}+\rho_{c}I_{c,k}+\sigma_{n}^{2}} \geq r_{c,k}-1,\,\forall k\in\mathcal{K},\\ 
& \frac{\rho_k a_{p,k}}{\sum_{i=1}^{K}\rho_i b_{ki}^{p}+\rho_{c}I_{c,k}+\sigma_{n}^{2}} \geq r_{p,k}-1,\,\forall k\in\mathcal{K},\\ 
&\;\;\;\;\;\mathbf{c} \geq \mathbf{0},\\ 
& \rho_{c} + \sum_{i=1}^{K} \rho_{i} \leq \rho_{\textrm{dL}}.
\end{align}
\end{subequations}
Here, (\ref{eq:MaxMin_2}) aims to maximize the lower bound of the objective function (\ref{eq:MaxMin_1}a) under the constraints (\ref{eq:MaxMin_1}b)-(\ref{eq:MaxMin_1}d). The equivalence between (\ref{eq:MaxMin_1}) and (\ref{eq:MaxMin_2}) is established based on the fact that at optimum, equality holds for constraints (\ref{eq:MaxMin_2}b)--(\ref{eq:MaxMin_2}g) and (\ref{eq:MaxMin_2}i). Next, we deal with the non-convex constraints (\ref{eq:MaxMin_2}f) and (\ref{eq:MaxMin_2}g) by further introducing auxiliary variables $\boldsymbol{\chi}_{c}=\{\chi_{c,1},\ldots,\chi_{c,K}\}$ and $\boldsymbol{\chi}_{p}=\{\chi_{p,1},\ldots,\chi_{p,K}\}$ representing the noise plus interference experienced by the common stream at UE sides and private streams, respectively. In addition, we introduce a variable vector $\boldsymbol{\nu}=\{\nu_{c},\nu_{1},\ldots,\nu_{K}\}$ such that $\nu_{c}^2=\rho_{c}$ and $\nu_{k}^{2}=\rho_{k},\,\forall k\in \mathcal{K}$. Consequently, constraint (\ref{eq:MaxMin_2}f) can be equivalently written as
\begin{subequations}\label{eq:MaxMin_3}
\begin{align}
\frac{\nu_{c}^{2}a_{c,k}}{\chi_{c,k}}&\geq r_{c,k}-1,\,\forall k\in\mathcal{K},\\
\chi_{c,k}&\geq \sum_{i=1}^{K}\nu_{i}^2 b_{ki}^{c}+\nu_{c}^{2}I_{c,k}+\sigma_{n}^{2},\,\forall k\in\mathcal{K}.
\end{align}
\end{subequations}
Similarly, for the private streams, constraint (\ref{eq:MaxMin_2}g) can be equivalently written as 
\begin{subequations}\label{eq:MaxMin_4}
\begin{align}
\frac{\nu_{k}^{2}a_{p,k}}{\chi_{p,k}}&\geq r_{p,k}-1,\,\forall k\in\mathcal{K},\\
\chi_{p,k}&\geq \sum_{i=1}^{K}\nu_{i}^2 b_{ki}^{p}+\nu_{c}^{2}I_{c,k}+\sigma_{n}^{2},\,\forall k\in\mathcal{K}.
\end{align}
\end{subequations}
The transmit power constraint, i.e., (\ref{eq:MaxMin_2}i) can be written as
\begin{equation}\label{eq:TxPw}
\nu_{c}^{2} + \sum_{i=1}^{K} \nu_{i}^{2} \leq \rho_{\textrm{dL}}.
\end{equation}
Therefore, problem (\ref{eq:MaxMin_1}) is equivalently transformed into
\begin{equation}\label{eq:MaxMin_5}
\begin{split}
    \max_{\substack{\boldsymbol{\nu},\mathbf{c},\boldsymbol{\alpha}_{p},\boldsymbol{\alpha}_{c},\\\mathbf{r}_{p},\mathbf{r}_{c},\boldsymbol{\chi}_{p},\boldsymbol{\chi}_{c},t}}&\;\;\;\;t\\
    s.t.&\;\;(\ref{eq:MaxMin_2}b), (\ref{eq:MaxMin_2}c),(\ref{eq:MaxMin_2}d), (\ref{eq:MaxMin_2}e),(\ref{eq:MaxMin_2}h),\\
    &\;\;(\ref{eq:MaxMin_3}a),(\ref{eq:MaxMin_3}b),(\ref{eq:MaxMin_4}a),(\ref{eq:MaxMin_4}b),(\ref{eq:TxPw}).
\end{split}
\end{equation}
The constraints of the transformed problem (\ref{eq:MaxMin_5}) are convex, with the exception of (\ref{eq:MaxMin_3}a) and (\ref{eq:MaxMin_4}a). To deal with these non-convex constraints, we use the first-order Taylor expansion to linearly approximate the non-convex parts of these constraints. Linear approximation of the left side of constraint (\ref{eq:MaxMin_3}a) around the point $(\nu_{c}^{[n]},\,\chi_{c,k}^{[n]})$ is given by 
\begin{equation}\label{eq:C_approx}
\frac{\nu_{c}^{2}a_{c,k}}{\chi_{c,k}}\geq a_{c,k}\Big(\frac{2\nu_{c}^{[n]}}{\chi_{c,k}^{[n]}}\nu_{c}-\frac{(\nu_{c}^{[n]})^{2}}{(\chi_{c,k}^{[n]})^{2}}\chi_{c,k}\Big)\triangleq \Psi_{c,k}^{[n]}(\nu_{c},\chi_{c,k}),    
\end{equation}
where $(\nu_{c}^{[n]},\chi_{c,k}^{[n]})$ are the values of variables $(\nu_{c},\chi_{c,k})$ in the $n^{th}$ iteration. Similarly, the non-convex part of constraint (\ref{eq:MaxMin_4}a), i.e., the left-hand side is linearly approximated around the point $(\nu_{k}^{[n]},\chi_{p,k}^{[n]})$, and the approximation is given by
\begin{equation}\label{eq:P_approx}
\frac{\nu_{k}^{2}a_{p,k}}{\chi_{p,k}}\geq a_{p,k}\Big(\frac{2\nu_{k}^{[n]}}{\chi_{p,k}^{[n]}}\nu_{k}-\frac{(\nu_{k}^{[n]})^{2}}{(\chi_{p,k}^{[n]})^{2}}\chi_{p,k}\Big)\triangleq \Psi_{p,k}^{[n]}(\nu_{k},\chi_{p,k}). 
\end{equation}
Based on the approximations in (\ref{eq:C_approx}) and (\ref{eq:P_approx}), at iteration $n$ problem (\ref{eq:MaxMin_1}) is approximated as,
\begin{equation}\label{eq:MaxMin_6}
\begin{split}
    \max_{\substack{\boldsymbol{\nu},\mathbf{c},\boldsymbol{\alpha}_{p},\boldsymbol{\alpha}_{c},\\\mathbf{r}_{p},\mathbf{r}_{c},\boldsymbol{\chi}_{p},\boldsymbol{\chi}_{c},t}}&\;\;\;\;t\\
    s.t.&\;\;\Psi_{c,k}^{[n]}(\nu_{c},\chi_{c,k})\geq r_{c,k}-1,\,\forall k\in \mathcal{K},\\
    &\;\; \Psi_{p,k}^{[n]}(\nu_{k},\chi_{p,k})\geq r_{p,k}-1,\,\forall k\in \mathcal{K},\\
    &\;\;(\ref{eq:MaxMin_2}b), (\ref{eq:MaxMin_2}c),(\ref{eq:MaxMin_2}d), (\ref{eq:MaxMin_2}e),(\ref{eq:MaxMin_2}h),\\
    &\;\; (\ref{eq:MaxMin_3}b), (\ref{eq:MaxMin_4}b), (\ref{eq:TxPw}).
\end{split}
\end{equation}
Problem (\ref{eq:MaxMin_6}) is convex \cite{Boyd@2004}, which can be solved using standard convex optimization algorithms, e.g., interior-point methods. In the numerical results section, we make use of the CVX toolbox in Matlab to solve problem (\ref{eq:MaxMin_6}). Algorithm \ref{alg:Max_MinA} outlines the SCA-based MaxMin power allocation algorithm of RS. In any iteration $n$, using the values $\boldsymbol{\nu}^{[n-1]}, \boldsymbol{\chi}_{p}^{[n-1]}, \boldsymbol{\chi}_{c}^{[n-1]}$ from the output of iteration $n-1$, problem (\ref{eq:MaxMin_6}) is solved and $t^{[n]}$, $\boldsymbol{\nu}^{[n]}, \boldsymbol{\chi}_{p}^{[n]}, \boldsymbol{\chi}_{c}^{[n]}$ are updated using the respective optimized values. The iterations continue till convergence is reached with a tolerance value $\epsilon$.  
\begin{algorithm}
\caption{MaxMin: SCA Algorithm}\label{alg:Max_MinA}
\begin{algorithmic}[1]
\State $\mathbf{Initialize}\:\: n\leftarrow 0,\: t^{[n]}\leftarrow 0, \boldsymbol{\nu}^{[n]}, \boldsymbol{\chi}_{p}^{[n]}, \boldsymbol{\chi}_{c}^{[n]}$
\State $\mathbf{Iterate}$
\State $\;\;n\leftarrow n+1;$
\State $\;\;Solve\;(\ref{eq:MaxMin_6})\;\textrm{using}$
\item[]$\;\;\;\;\boldsymbol{\nu}^{[n-1]}, \boldsymbol{\chi}_{p}^{[n-1]}, \boldsymbol{\chi}_{c}^{[n-1]} \textrm{and}\;\textrm{denote}\;\textrm{optimal}\;\textrm{values}\;\textrm{of}\;t,$
\item[]$\;\;\;\;\boldsymbol{\nu},\,\boldsymbol{\chi}_{p},\,\boldsymbol{\chi}_{c} \;\;\;\textrm{as}\;\;\; t^{*},\,\boldsymbol{\nu}^{*},\, \boldsymbol{\chi}_{p}^{*},\,\boldsymbol{\chi}_{c}^{*}.$
\State $\mathbf{Update}\;t^{[n]}\leftarrow t^{*},\,\boldsymbol{\nu}^{[n]}\leftarrow\boldsymbol{\nu}^{*},\,\boldsymbol{\chi}_{p}^{[n]}\leftarrow \boldsymbol{\chi}_{p}^{*},\,\boldsymbol{\chi}_{c}^{[n]}\leftarrow \boldsymbol{\chi}_{c}^{*}$
\State $\mathbf{Until}\;\;|t^{[n]}-t^{[n-1]}|< \epsilon $
\end{algorithmic}
\end{algorithm}
\par \textbf{Initialization}: Since the variable $\boldsymbol{\nu}$ and auxiliary variables $\boldsymbol{\chi}_{c}$ and $\boldsymbol{\chi}_{p}$ all depend on $\boldsymbol{\rho}$, we begin with describing the initialization of $\boldsymbol{\rho}$. The initial power allocation is done by finding a feasible point $\boldsymbol{\rho}^{[0]}$ satisfying the transmit power constraint in (\ref{eq:MaxMin_1}c). With $\zeta\in[0,1]$ denoting the fraction of total power allocated to the common stream, we have $\rho_c=\zeta\rho_{\textrm{dL}}$. Furthermore,  $\rho_{k}=(1-\zeta){\rho_{\textrm{dL}}}/{K},\,\forall k\in\mathcal{K}$ is the initial power allocated to each private stream. Consequently, ${\nu}_{c}^{[0]}$ and ${\nu}_{k}^{[0]}$ are respectively initialized as $\nu_{c}^{[0]}=\sqrt{{\rho}_{c}^{[0]}}$ and ${\nu}_{k}^{[0]}=\sqrt{{\rho}_{k}^{[0]}},\,\forall k\in\mathcal{K}$, satisfying the transmit power constraint in (\ref{eq:TxPw}). We initialize $\boldsymbol{\chi}_{c}$ and $\boldsymbol{\chi}_{p}$ by replacing the inequalities with equalities in (\ref{eq:MaxMin_3}b) and (\ref{eq:MaxMin_4}b), respectively.
\par \textbf{Convergence}: Since the constraints (\ref{eq:MaxMin_3}a) and (\ref{eq:MaxMin_4}a) are relaxed by the first-order lower bounds (\ref{eq:C_approx}) and (\ref{eq:P_approx}), respectively, a feasible solution of problem (\ref{eq:MaxMin_6}) at iteration $n$ is also a feasible solution at iteration $n+1$. Therefore, the optimized value of $t$ is always  non-decreasing with $n$. As $t$ is bounded by the transmit power constraint in (\ref{eq:TxPw}), Algorithm \ref{alg:Max_MinA} is guaranteed to converge. However, due to linear approximations (\ref{eq:C_approx}) and (\ref{eq:P_approx}), the algorithm is not guaranteed to converge to the global optimum.  Initialization plays an important role in determining the optimized value of $t$ and therefore, solving problem (\ref{eq:MaxMin_6}) for different values of $\zeta$ and selecting the optimal power allocation that maximizes $t$ helps us achieve significantly better performance. 
\par {\textbf{Complexity}: Algorithm \ref{alg:Max_MinA} involves optimization of $8K+2$ variables and $9K+1$ constraints and therefore has the computational complexity of the order of \cite{Boyd@2004}
\begin{equation}\label{eq:maxmin_complexity}
    \mathcal{O}\left(N_{o}\max\left\{4\left(4K+1\right)^{2}\left(9K+1\right),\,F_{2}\right\}\right),
\end{equation}
where $F_{2}$ is the cost of evaluating the first and second
derivatives of the objective and constraint functions in (\ref{eq:MaxMin_6}). $N_{o}$ is the number of iterations required for Algorithm \ref{alg:Max_MinA} to converge to an optimal (local or global) point. Note that although the joint optimization of $\boldsymbol{\rho}$, $\mathbf{c}$ and $t$ noticeably increases the complexity of each iteration in Algorithm \ref{alg:Max_MinA} as the number of UEs increase, careful initialization of power allocation coefficients can help in hastening the convergence of the SCA algorithm to a few iterations.}
\begin{remark}
{We would like to highlight that the SCA method is opted for MaxMin because of the joint optimization of power allocation coefficients and share of the common SE allocated to each UE in order to maximize the minimum SE. MaxSum-SE and MaxSINR power allocation problems can also be formulated and solved using the SCA method. However, we choose and illustrate Algorithm \ref{alg:MaxSumSE_alg} for MaxSum-SE because different from MaxMin, MaxSum-SE only optimizes power allocation coefficients without the optimization of coupled common SE allocation. It is therefore possible to develop a power allocation strategy with a lower complexity and still achieve good SE performance. Since Algorithm \ref{alg:MaxSumSE_alg} has a low time-complexity and builds on the popular uniform power allocation scheme of NoRS in practical networks, we choose Algorithm \ref{alg:MaxSumSE_alg} to maximize the sum-SE of RS. For MaxSINR, we opt for Algorithm \ref{alg:MaxSINR_alg} because it provides the optimal solution to problem (\ref{eq:MaxSINR_1}), as opposed to SCA, which gives a sub-optimal solution. Nevertheless, Appendix \ref{App_B} demonstrates the use of SCA algorithm to solve MaxSum-SE and MaxSINR problems of RS, and their comparison with Algorithm \ref{alg:MaxSumSE_alg} and Algorithm \ref{alg:MaxSINR_alg}, respectively.}
\end{remark}

\section{Numerical Results}\label{NumRes}
\begin{figure}
\centering
\subfloat[Rectangular Topology]{\includegraphics[width=4.6cm,height=4.7cm]{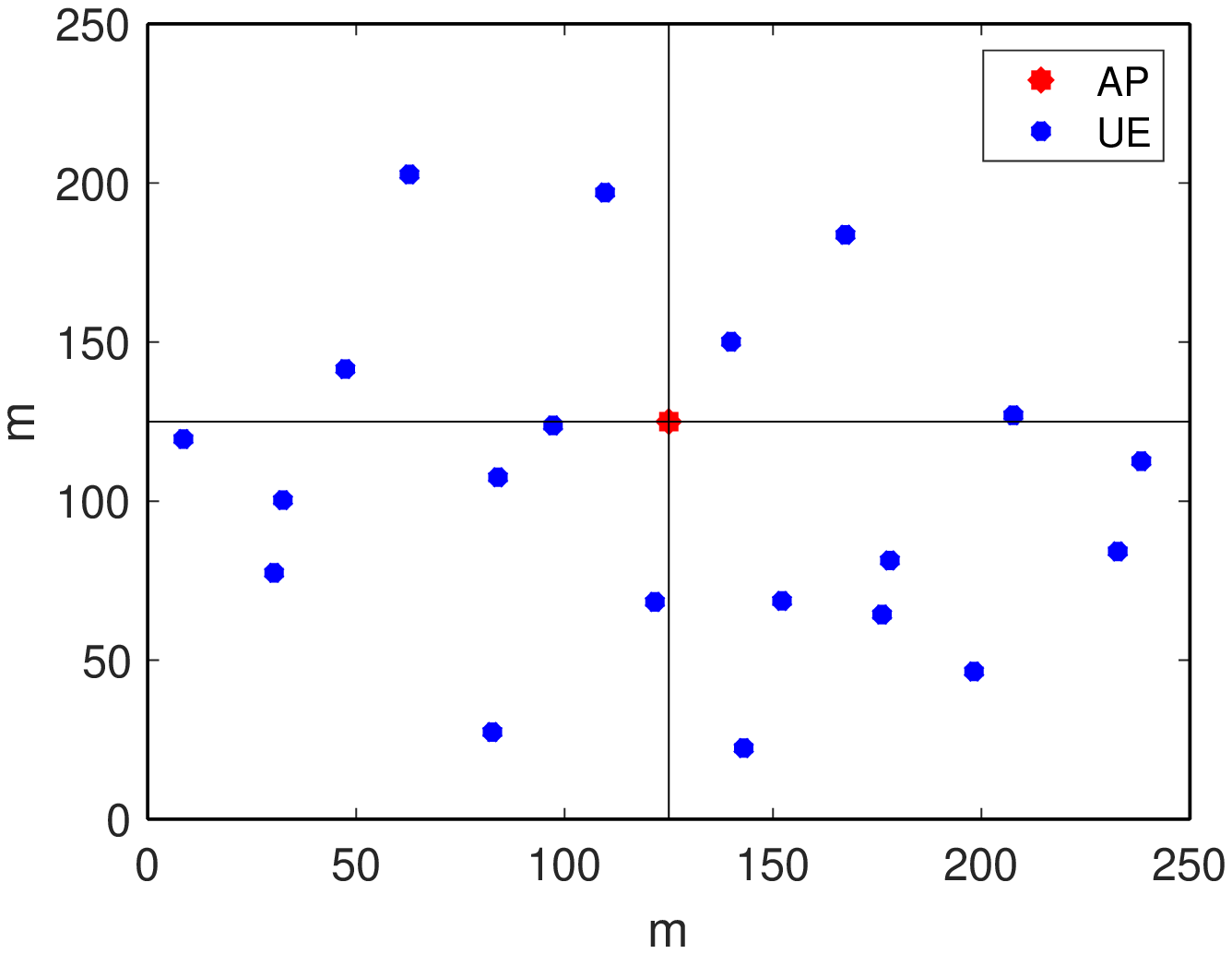}}
\subfloat[Circular Topology]{\includegraphics[width=4.6cm,height=4.7cm]{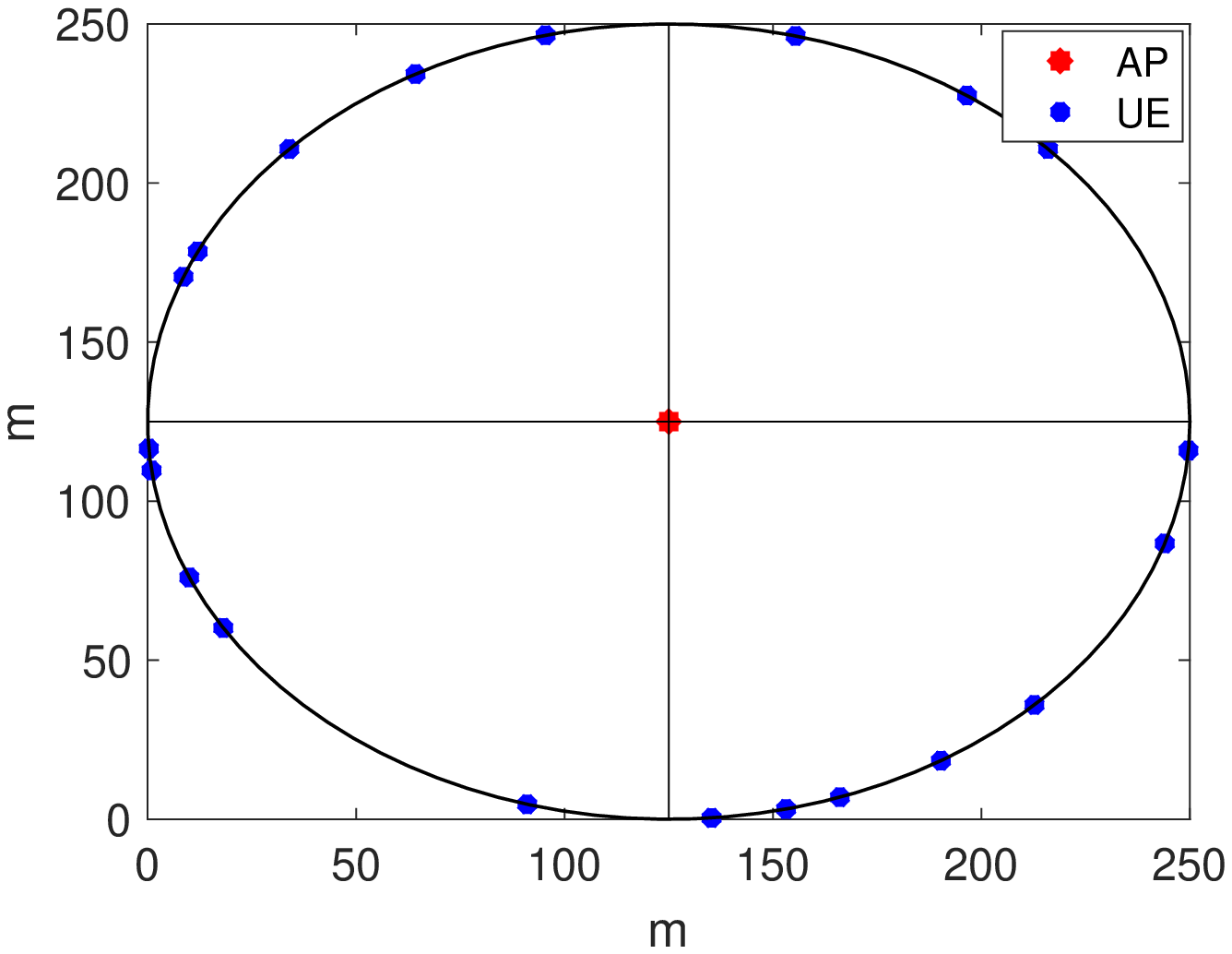}}
\caption{Network topologies.}%
\label{fig:Topology_Diag}
\end{figure}

To quantitatively compute the SE achieved by RS in TDD MaMIMO, we consider different topologies to capture different use cases and obtain the SE performance of RS and NoRS transmission strategies. For each topology, the SE results are obtained by averaging the SE over $100$ setups. In each setup, the lower bound of the SE performance is calculated considering $200$ coherence intervals. The two topologies considered in this paper are: 1) a rectangular topology where UEs are randomly distributed within a rectangular area of size $250\times 250\,\textrm{m}^{2}$, illustrated in Fig.~\ref{fig:Topology_Diag}(a); and 2) a circular topology where UEs are randomly distributed around a circle of radius $r=125\,\textrm{m}$, illustrated in Fig.~\ref{fig:Topology_Diag}(b). {Here, both topologies are considered to appropriately capture use cases like conventional cellular use case, ultra-reliable low-latency communications (URLLC), mMTC, and crowded scenarios}.       
\par At the transmitter side, the BS antennas are placed in a uniform linear array with half-wavelength antenna spacing. The large scale fading parameter path loss $\beta_{k}$ for UE $k$ is modelled in dB 
as \cite[eq (2.3)]{massivemimobook} 
\begin{equation}\label{eq:Pathloss}
\beta_{k}=\Gamma - 10\eta\log_{10}(\frac{d_{k}}{1\textrm{km}})+{S}_{k},
\end{equation}
where $d_{k}$ is the distance between the BS and UE $k$ in $\textrm{km}$, the pathloss exponent $\eta$ determines how fast the signal power decays and $\Gamma$ is the channel gain at a distance $1\,\textrm{km}$. {${S}_{k}\in \mathcal{N}(0,\sigma_{s}^{2})$} is the shadow fading  coefficient. Subsequently, the corresponding channel correlation matrix for each UE is generated by assuming that each channel consists of $S=10$ clusters following the Gaussian scattering model in \cite[Sec 2.6]{massivemimobook}. Hence, the $(m_{1},m_{2})\textrm{th}$ element of correlation matrix of UE $k$ is given by
\begin{equation}
\begin{split}
    [\mathbf{R}_{k}]_{m_{1},m_{2}}&=\beta_{k}\times\frac{1}{S}\times\\
    &\sum_{s=1}^{S}e^{\textrm{i}\pi(m_{1}-m_{2})\sin(\varphi_{k,s})}e^{-\frac{\sigma_{\varphi}^{2}}{2}\big(\pi(m_{1}-m_{2})\cos(\varphi_{k,s})\big)^{2}}.
\end{split}
\end{equation}
Let $\varphi_{k}$ be the geographical angle to UE $k$ from the BS. Cluster $s$ is characterized by a randomly generated nominal angle-of-arrival $\varphi_{k,s}\sim \mathcal{U}\{\varphi_{k}-40^{\degree},\varphi_{k}+40^{\degree}\}$ and multipath components have their angles distributed around the corresponding nominal angle with variance $\sigma_{\varphi}=15^{\degree}$. {For uncorrelated Rayleigh fading, the spatial correlation matrix simply boils down to $\mathbf{R}_{k}=\beta_{k}\mathbf{I}_{M}$}\cite{Luca@MaMIMO2}. The simulation parameters are reported in Table \ref{tab:Para_table}. 
\begin{table}
	\caption{Simulation parameters}
	    \label{tab:Para_table}\centering
	\begin{tabular}{l l}
		\toprule[0.4mm]
		\textbf{Parameter} & \textbf{Value}\\
   Path Loss coefficients	& $\Gamma=-148.1$, $\eta=3.76$, $\sigma_{s}^{2}=16$\\
   TDD parameters (samples) & {$\tau = 200,\,\tau_{p}=20,\,\tau_{d}= 190$}\\
   Total transmit powers &  $\rho_{\rm{ul}} = 10\,\textrm{dBm}$, $\rho_{\textrm{dL}}= 20\,\textrm{dBm}$\\
   Noise powers & $\sigma_{\rm{ul}}^{2}=\sigma_{n}^{2} = -94\,\textrm{dBm}$\\
		\bottomrule[0.4mm]
	\end{tabular}
\end{table}
\par We consider NoRS (a conventional SDMA-based MaMIMO strategy) as the baseline strategy to compare the SE performance of RS.\footnote{{We do not consider PD-NOMA for SE performance comparison as it is unable to fully exploit the spatial domain in multi-antenna settings\cite{Bruno@NOMA}. RSMA has been shown to outperform PD-NOMA in both DoF and SE performance in multi-user MIMO networks\cite{Bruno@NOMA}. Moreover, DL training is required for PD-NOMA. To that end, \cite{emil@NOMA} compares NoRS with a PD-NOMA strategy in TDD MaMIMO and showed that NoRS significantly outperforms PD-NOMA in terms of SE performance, even with DL training utilized for UEs in PD-NOMA. Therefore, in this paper, we restrict the SE performance comparison of RS with the NoRS strategy only.}} {For a fair SE performance comparison, we choose MR precoding for UEs and calculate the SE performance of NoRS for all three different power allocation schemes as following,} 
\begin{itemize}
    \item \textit{MaxSum-SE}: For MaxSum-SE, baseline NoRS results are obtained by switching off the common stream, i.e., by computing the sum-SE performance for $\zeta=1$ in Algorithm \ref{alg:MaxSumSE_alg}.
    \item \textit{MaxSINR}: MaxSINR problem of NoRS is formulated following equation $(7.8)$ in \cite{massivemimobook}. MaxSINR problem with NoRS is also a GP problem and is solved using CVX.
    \item \textit{MaxMin}: We opt for two MaxMin algorithms to obtain the SE performance of the NoRS strategy. In the first method, the common stream is switched off by allocating zero power, i.e., forcing $\rho_{c}=0$, and then solving problem (\ref{eq:MaxMin_6}) using the SCA method. The second method is solving the MaxMin fairness problem $(7.7)$ formulated in \cite{massivemimobook}, which gives a globally optimal solution for a NoRS strategy by employing the Bisection algorithm. {The first method (SCA) is considered to ensure fairness in MaxMin SE performance comparison of the RS and NoRS strategies, whereas the second method allows us to compare a sub-optimal SCA-based power allocation scheme of NoRS with the globally optimal algorithm of NoRS based on Bisection.} The power allocation coefficients are initialized as $\rho_{c}=0.1\rho_{\textrm{dL}}$ and $\rho_{k}={0.9\rho_{\textrm{dL}}}/{K},\, \forall k\in\mathcal{K}$ for the MaxMin scheme of RS, whereas the initialization is $\rho_{k}={\rho_{\textrm{dL}}}/{K},\, \forall k\in\mathcal{K}$ for SCA-based MaxMin scheme of NoRS. 
\end{itemize}
\subsection{{MaxMin}}\label{numRes_Ortho}
{For RS, the SE per UE performance is obtained by employing Algorithm \ref{alg:Max_MinA} delineated in Section \ref{PowerAlloc}.C, whereas the MaxMin power allocation scheme for the baseline NoRS strategy is discussed in Section \ref{NumRes} (SCA and Bisection both).}  
  \subsubsection{{No Pilot Contamination}} 
\begin{figure}
\centering
{\includegraphics[width=6.5cm,height=5.5cm]{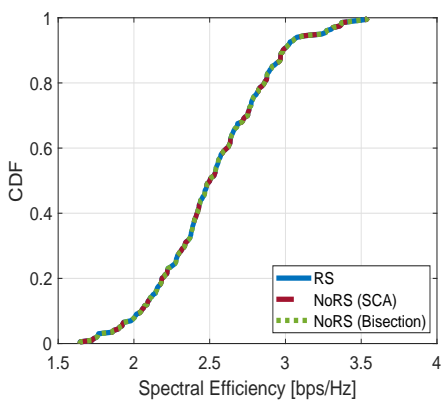}}
\caption{Average SE per UE, $M=100,\,K=8,$ orthogonal pilots.}%
\label{fig:EqualPA_Ortho}
\end{figure}

{To appreciate the role of RS in mitigating the deleterious effect of pilot contamination, we first analyze the SE per UE performance of both RS and NoRS transmission strategies when there is no pilot contamination, i.e., when all UEs use mutually orthogonal pilot sequences for UL training.} Fig.~\ref{fig:EqualPA_Ortho} illustrates the average SE per UE performance of both RS and NoRS strategies in the rectangular topology and spatially correlated UE channels. We observe that since there is no pilot contamination, {i.e., the CSIT quality is ``good"}, both transmission strategies achieve high SE performance. Moreover, RS has no gain over NoRS. {Note that since RS encapsulates a conventional linearly precoded multi-user MIMO strategy, under perfect CSIT conditions, RS will always achieve a SE that is better than or equal to NoRS for the same optimization technique\cite{Bruno@NOMA}}. As a result, with no pilot contamination, RS achieves the same SE performance as NoRS strategy, i.e., no power is allocated to the common streams, and all the transmit power is allocated to the private streams. The SE per UE result is similar for the circular topology and therefore is not illustrated for brevity. Furthermore, the SE per UE performance with spatially uncorrelated channels is consistent with the result of spatially correlated channels in the sense that RS and NoRS achieve the same performance, and therefore is not shown to avoid redundancy. 

\subsubsection{{Severe pilot contamination}}
\begin{figure}
\centering
\subfloat[$M=100$, $K=8$]{\includegraphics[width=4.6cm,height=4.7cm]{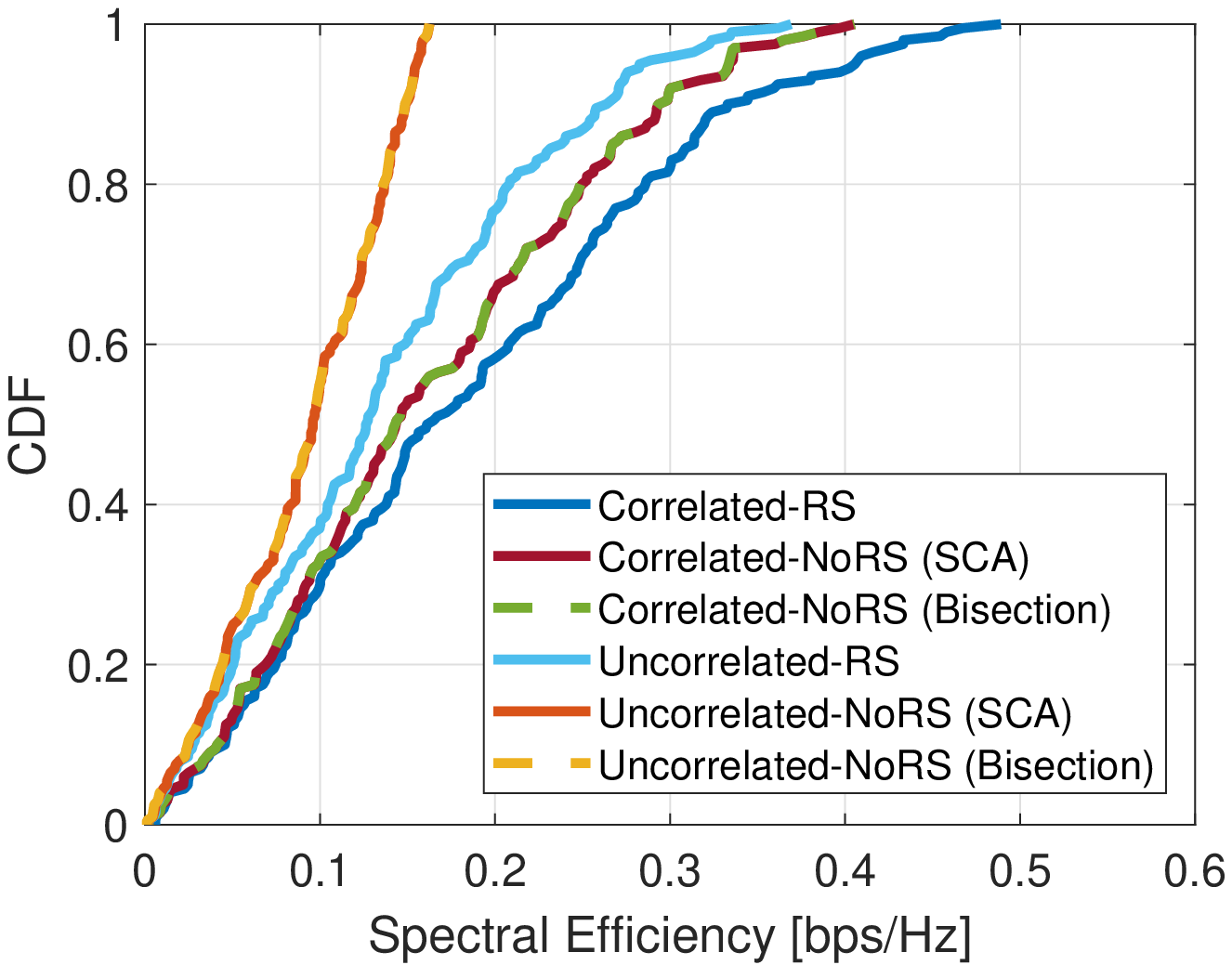}}
\subfloat[$M=100$]{\includegraphics[width=4.6cm,height=4.7cm]{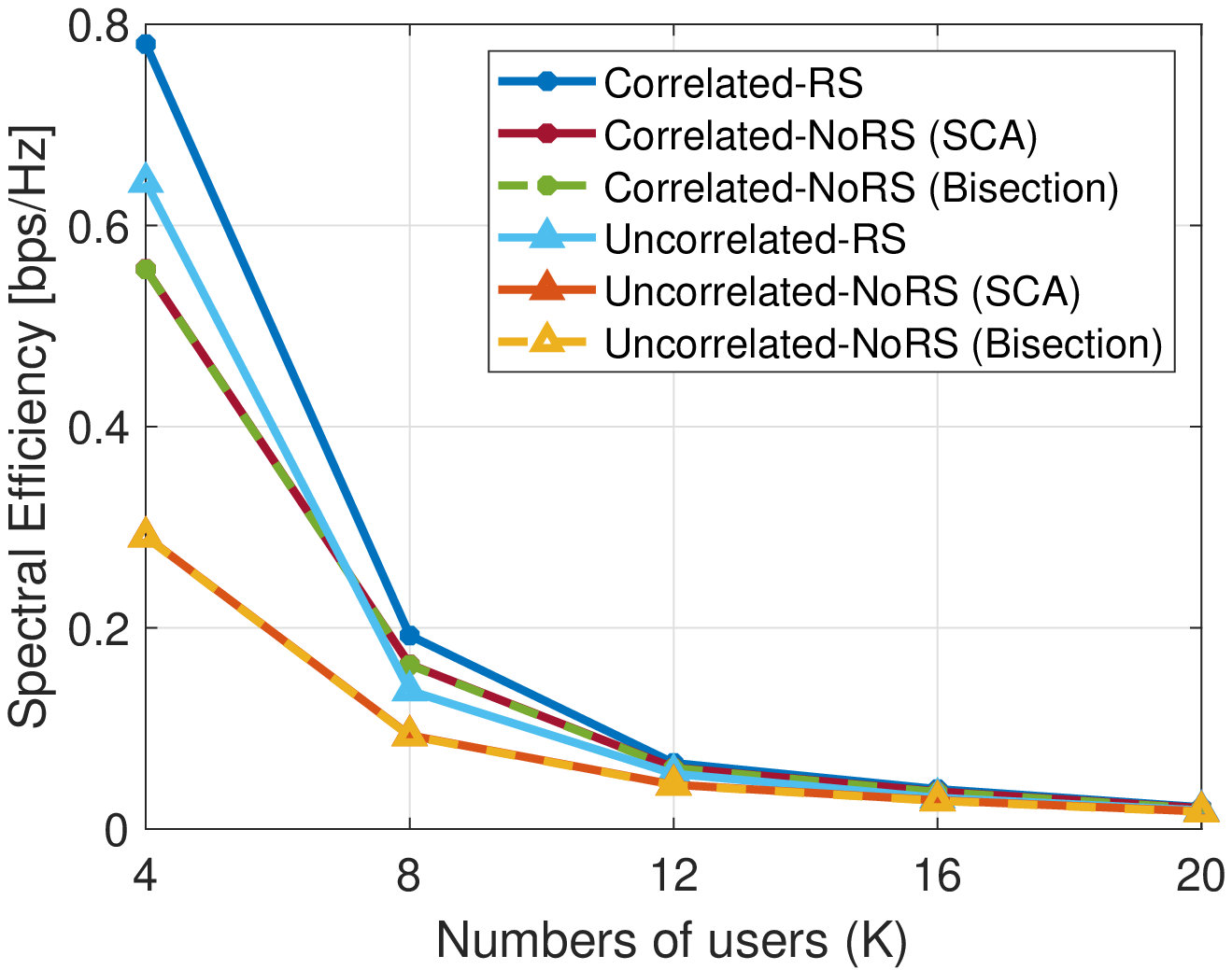}}
\caption{Average SE per UE in rectangular topology.}%
\label{fig:MaxMinPA_cellular}
\end{figure}

Next, we move to the other extreme and illustrate the SE per UE of both transmission strategies under severe pilot contamination for different topologies and channel propagation environments. Fig.~\ref{fig:MaxMinPA_cellular} illustrates the SE per UE performance of RS and NoRS strategies in the rectangular topology. From Fig.~\ref{fig:MaxMinPA_cellular}(a), we observe that the RS strategy can better manage the interference and achieves a better SE per UE performance than the NoRS strategy. Moreover, the gain of RS over NoRS is larger when the channel fading is uncorrelated. {The reason for a larger SE gain in the case of uncorrelated fading is that the channel estimates of two UEs utilizing the same pilot sequence become parallel vectors that only differ in scaling, i.e., $\widehat{\mathbf{g}}_{k}=({\beta_{k}}/{\beta_{i}})\widehat{\mathbf{g}}_{i},\,\forall i,\,k\in\mathcal{K}$. As a result, the BS is unable to separate UE channels that are identically distributed (up to a scaling factor), resulting in extremely poor quality and correlated channel estimates, which ultimately leads to a decrease in the SE performance of both the RS and NoRS strategy\cite{massivemimobook,emil@Luca}. However, due to the robustness of RS under imperfect CSIT, the performance loss in RS is significantly less severe compared to NoRS.} Fig.~\ref{fig:MaxMinPA_cellular}(b) shows the average SE per UE versus the number of UEs for the same network layout. For $K=4$, the relative SE gain\footnote{Relative SE gain is calculated as $\frac{\textrm{SE}^{\textrm{RS}}-\textrm{SE}^{\textrm{NoRS}}}{\textrm{SE}^{\textrm{RS}}}$.} of RS over NoRS is $28.6\%$ for correlated fading channels and $58.7\%$ for uncorrelated fading channel. As the number of UE increases, both RS and NoRS show a sharp SE loss which can be attributed to multiple reasons. First, since all UEs are using the same pilot, the effect of pilot contamination becomes more severe as the number of UEs increases. Second, maximizing the minimum SE becomes more difficult as the number of UEs increases. For RS, the sharp decline also stems from the additional constraint of maximizing the common SE as the common stream is to be decoded by all UEs. Therefore, the SE gain of RS over NoRS decreases as the number of UEs increases.\footnote{{As the number of UEs increases, more advanced RSMA schemes could be used in MaMIMO to improve the SE performance \cite{mao2017rate,Minbo2016MassiveMIMO}.}} 
\begin{figure}
\centering
\subfloat[$M=100$, $K=8$]{\includegraphics[width=4.6cm,height=4.7cm]{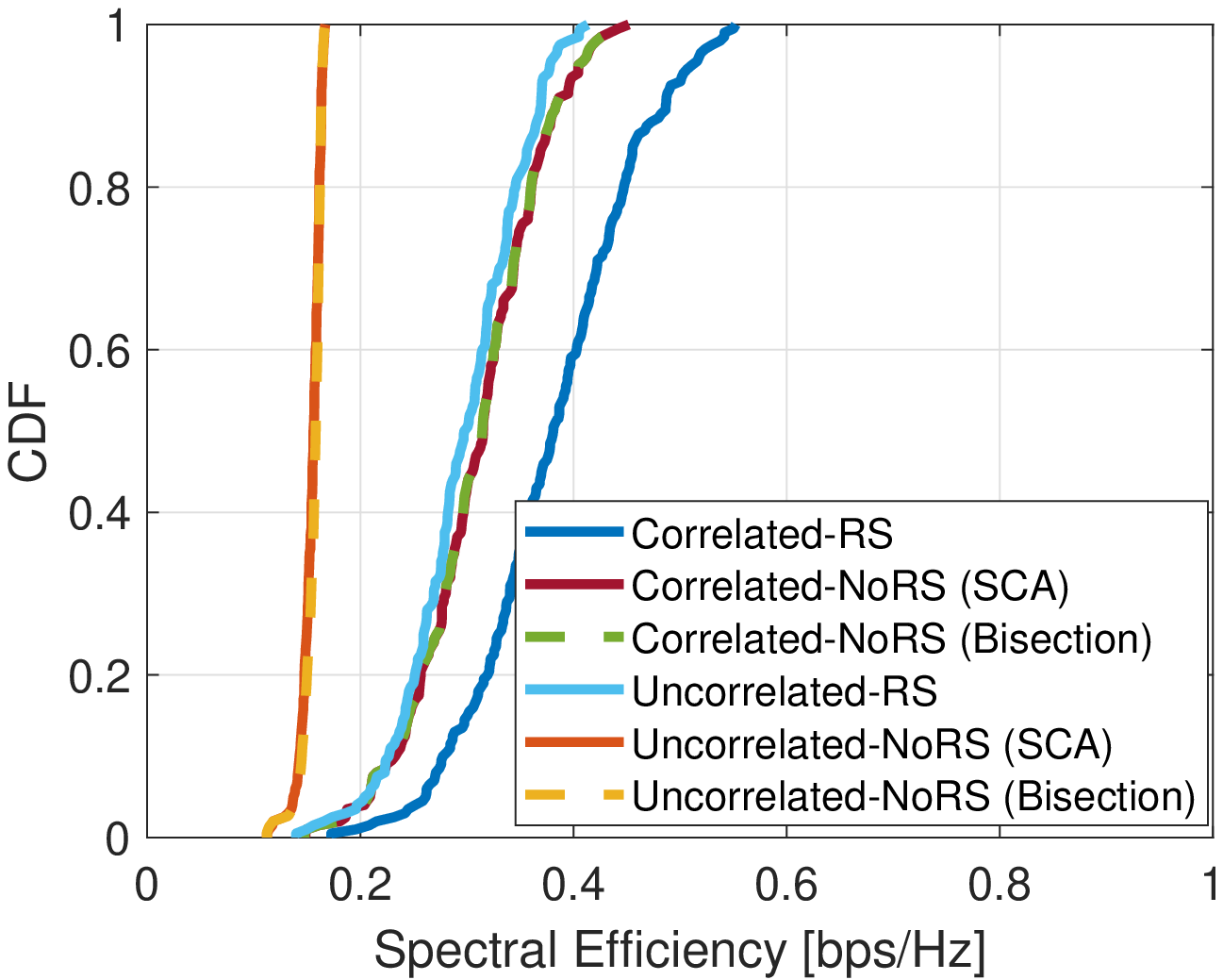}}
\subfloat[$M=100$]{\includegraphics[width=4.6cm,height=4.7cm]{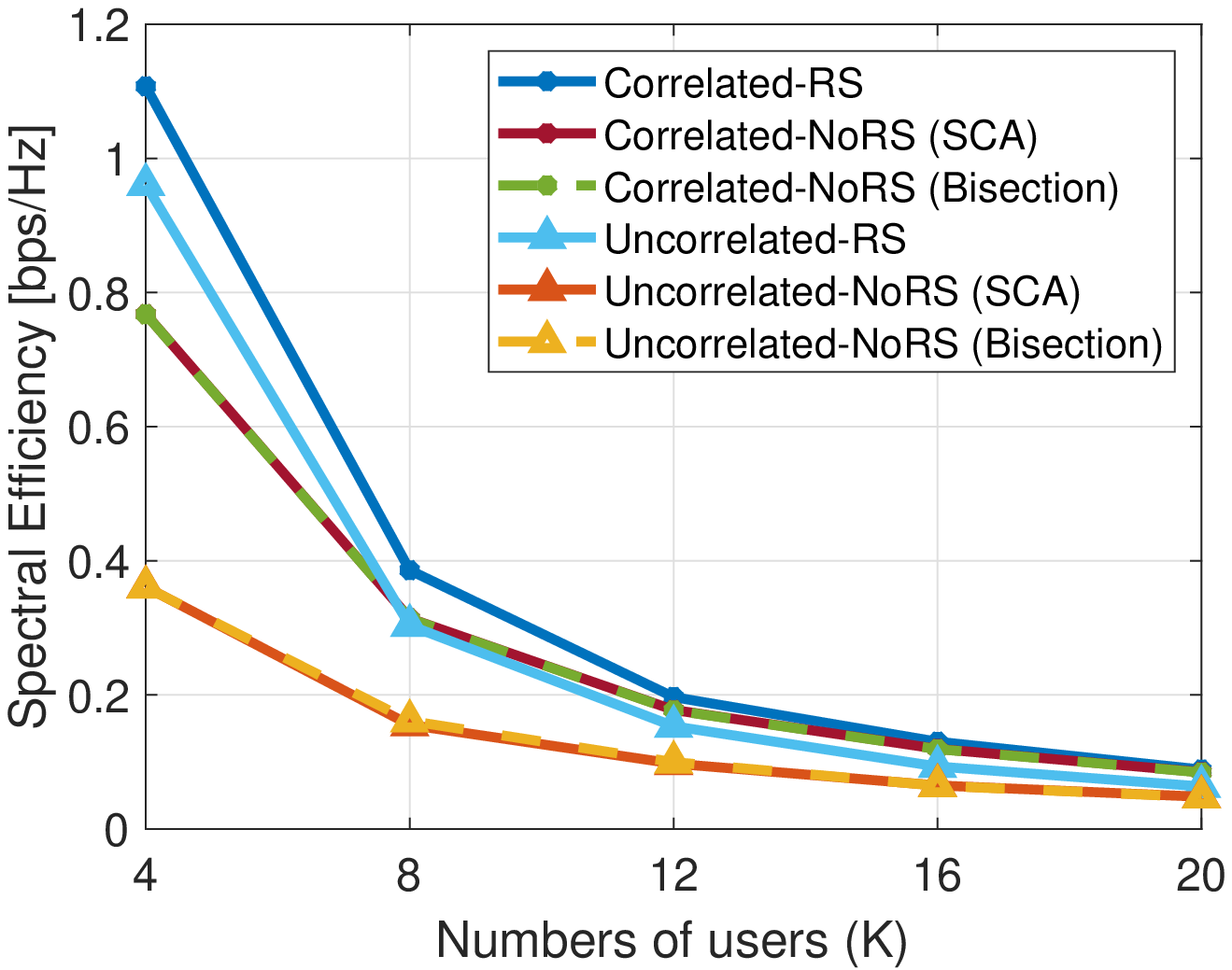}}
\caption{ Average SE per UE in circular topology.}%
\label{fig:MaxMinPA_Circular}
\end{figure}
\par Next, we look at the performance of the transmission strategies in the circular topology. From Fig.~\ref{fig:MaxMinPA_Circular}, we observe that the gain of RS over NoRS is more explicit in the circular topology compared to the rectangular topology. This accentuated gain is due to the ease of the constraint of maximizing the common SE as UEs experience similar path loss. In this scenario, for $K=4$, the gain of RS over NoRS is $30\%$ for correlated fading channels and $62\%$ for uncorrelated fading channels. Similarly, as observed in Fig.~\ref{fig:MaxMinPA_Circular}(b), the decline in average SE per UE is relatively less sharp in circular topology because equal path loss aids in maximizing both the common SE and minimum total SE of a UE. As a result, the SE performance of RS is significantly higher than NoRS, even for a higher number of UEs. This is particularly beneficial for fixed mMTC scenarios where multiple active UEs (devices) have a high probability of sharing a pilot sequence and typically have low SE requirements in both UL and DL. 
\par
\begin{figure}
\centering
\subfloat[Rectangular Topology, $\theta=\pi/4$]{\includegraphics[width=4.6cm,height=4.7cm]{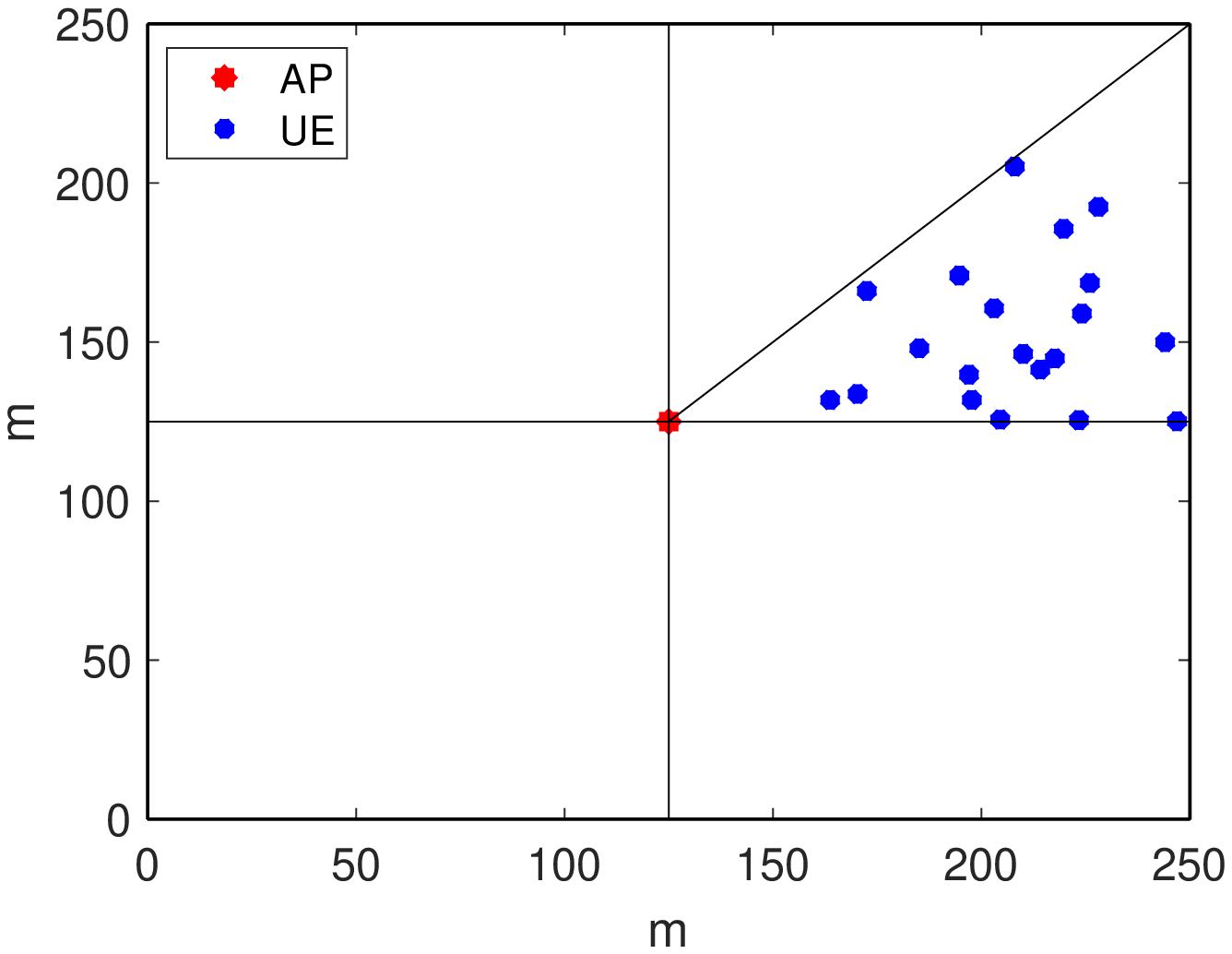}}
\subfloat[Circular Topology, $\theta=\pi/4$]{\includegraphics[width=4.6cm,height=4.7cm]{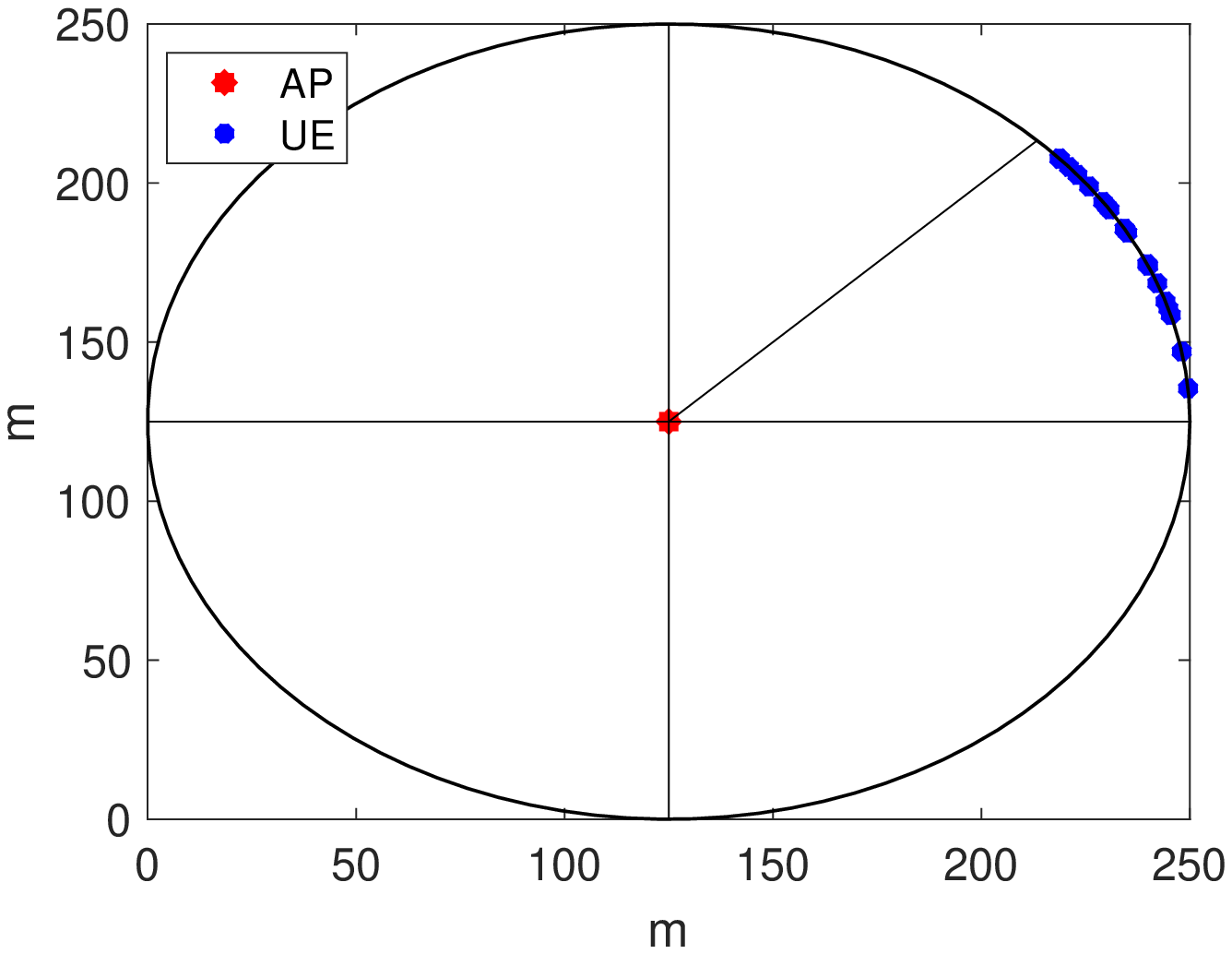}}
\caption{Network topologies with UEs confined within a sector.}%
\label{fig:Topology_Diag_Pi4}
\end{figure}

\begin{figure}
\centering
\subfloat[Rectangular]{\includegraphics[width=4.6cm,height=4.7cm]{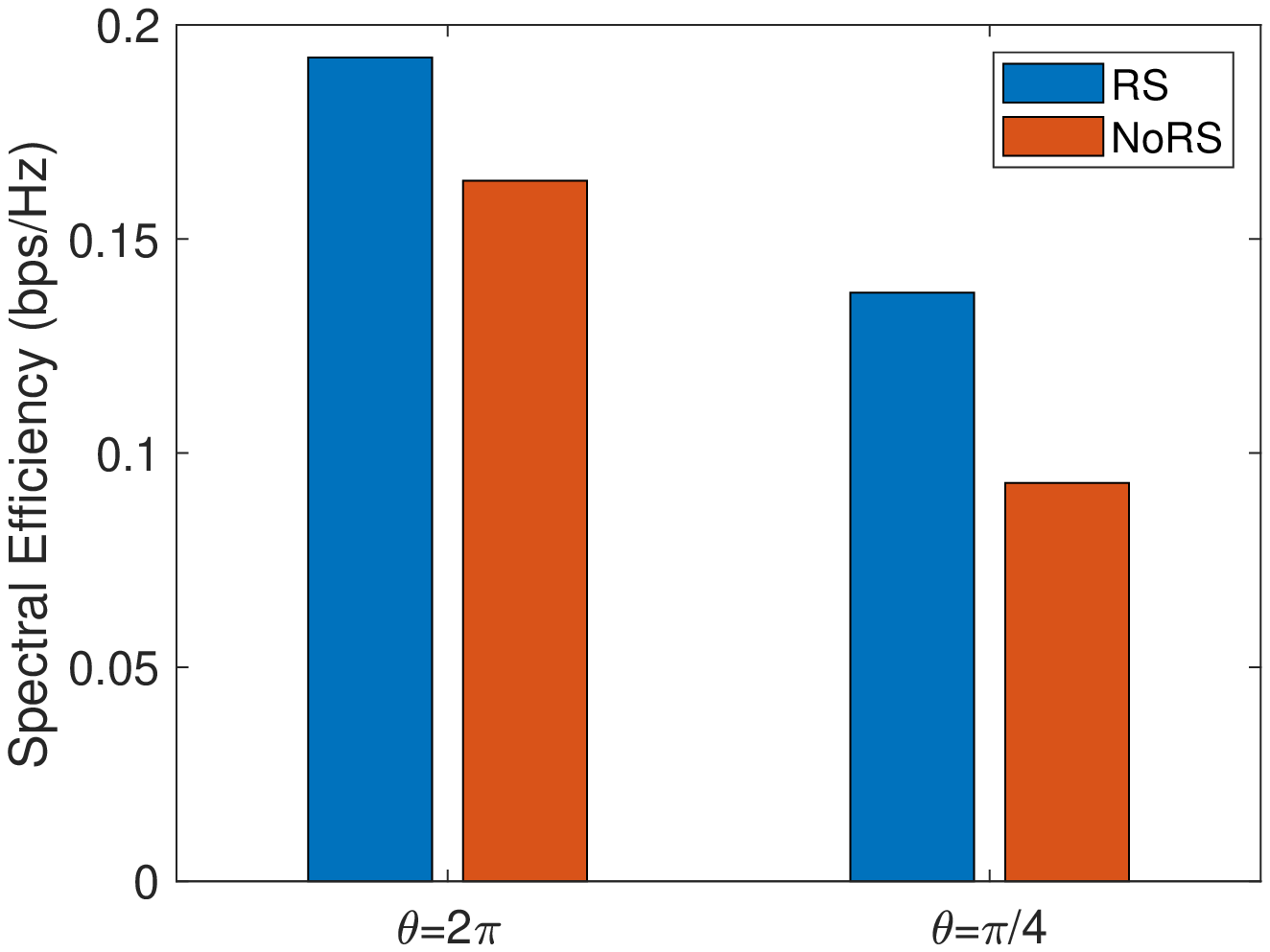}}
\subfloat[Circular]{\includegraphics[width=4.6cm,height=4.7cm]{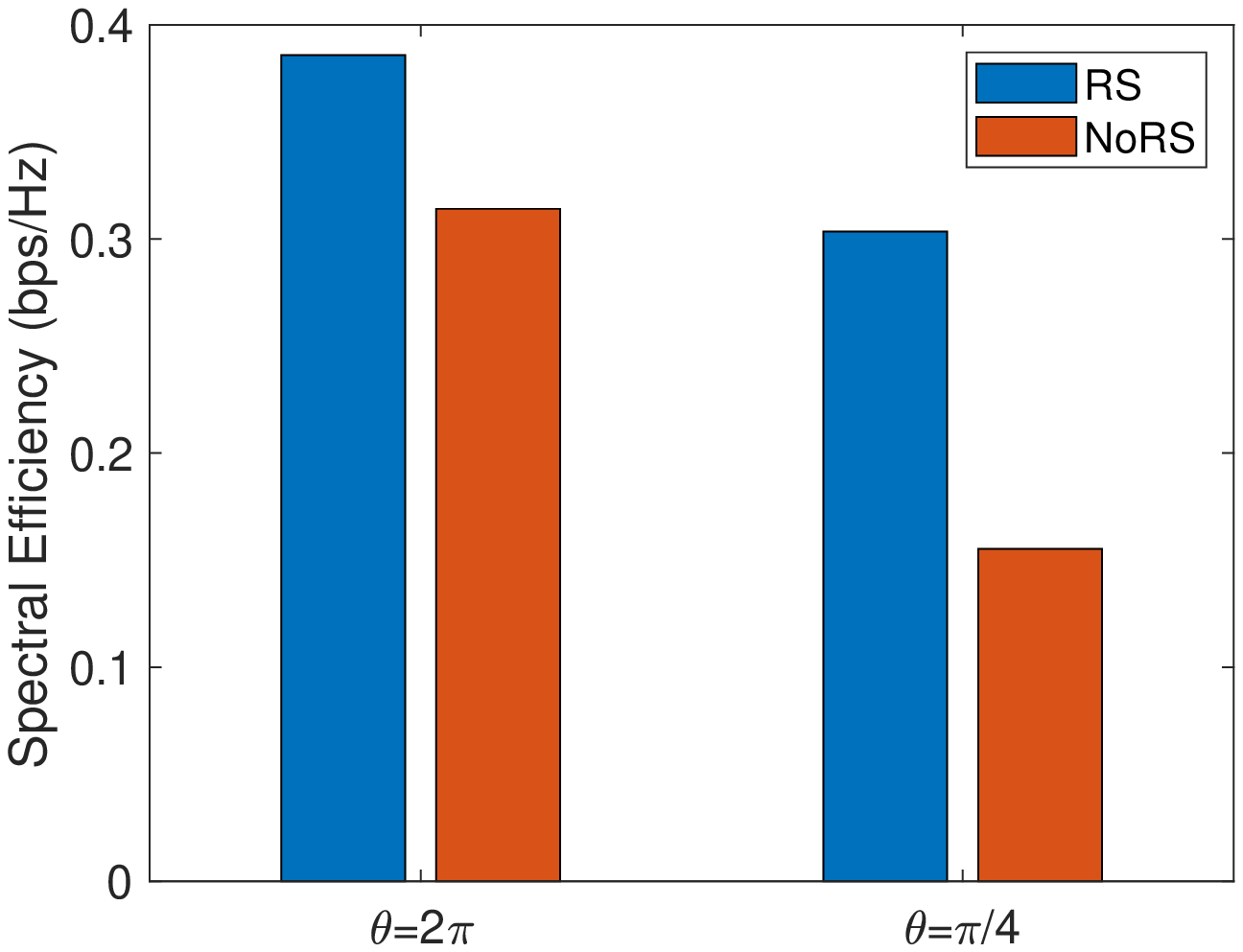}}
\caption{ Average SE per UE with spatially correlated channels for different network layouts, $M=100$ and $K=8$.}
\label{fig:MaxMinPA_Sector}
\end{figure}

 It should be highlighted that with spatially correlated channels, the relatively low gains of RS over NoRS for a higher number of UEs is not indicative of the true performance of RS. Since the UEs are randomly distributed within an area in the rectangular topology, and along the circumference in circular topology, the channels of UEs are highly distinguishable in the spatial domain. In such a scenario, the MMSE estimation is better placed to separate the channel estimates in the spatial domain and, even with a single pilot used for UL estimation, can better mitigate the effect of pilot contamination. However, if the UEs are packed within a sector, e.g. crowded scenarios, the efficacy of MMSE estimation in mitigating pilot contamination will suffer a SE loss. To elaborate, we compare the average SE per UE of RS and NoRS for the following two network layouts: 1) first, when the network layout is the same as Fig.~\ref{fig:Topology_Diag}, i.e., UEs are randomly distributed in the whole area of consideration ($\theta=2\pi$) and 2) when UEs are randomly distributed within a confined sector ($\theta=\pi/4$), as illustrated in Fig.~\ref{fig:Topology_Diag_Pi4}. It can be observed in Fig.~\ref{fig:MaxMinPA_Sector}(a) that though the achieved SE per UE decreases for both RS and NoRS as $\theta$ decreases to $\pi/4$ due to reduced CSIT quality, the relative SE gain of RS over NoRS in rectangular topology increases from $15\%$ to $23.5\%$. Similarly, the relative SE gain of RS over NoRS in circular topology illustrated in Fig.~\ref{fig:MaxMinPA_Sector}(b) increases from $18.6\%$ to $35.33\%$. These results further attest to the superiority of RS over NoRS strategy in different use cases by robustly managing interference and achieving high SE performance.
\vspace{-0.5cm}
\subsection{{MaxSum-SE and MaxSINR}}\label{lowcomplex_PA}
\begin{figure}
\centering
\subfloat[MaxSum-SE]{\includegraphics[width=4.6cm,height=4.7cm]{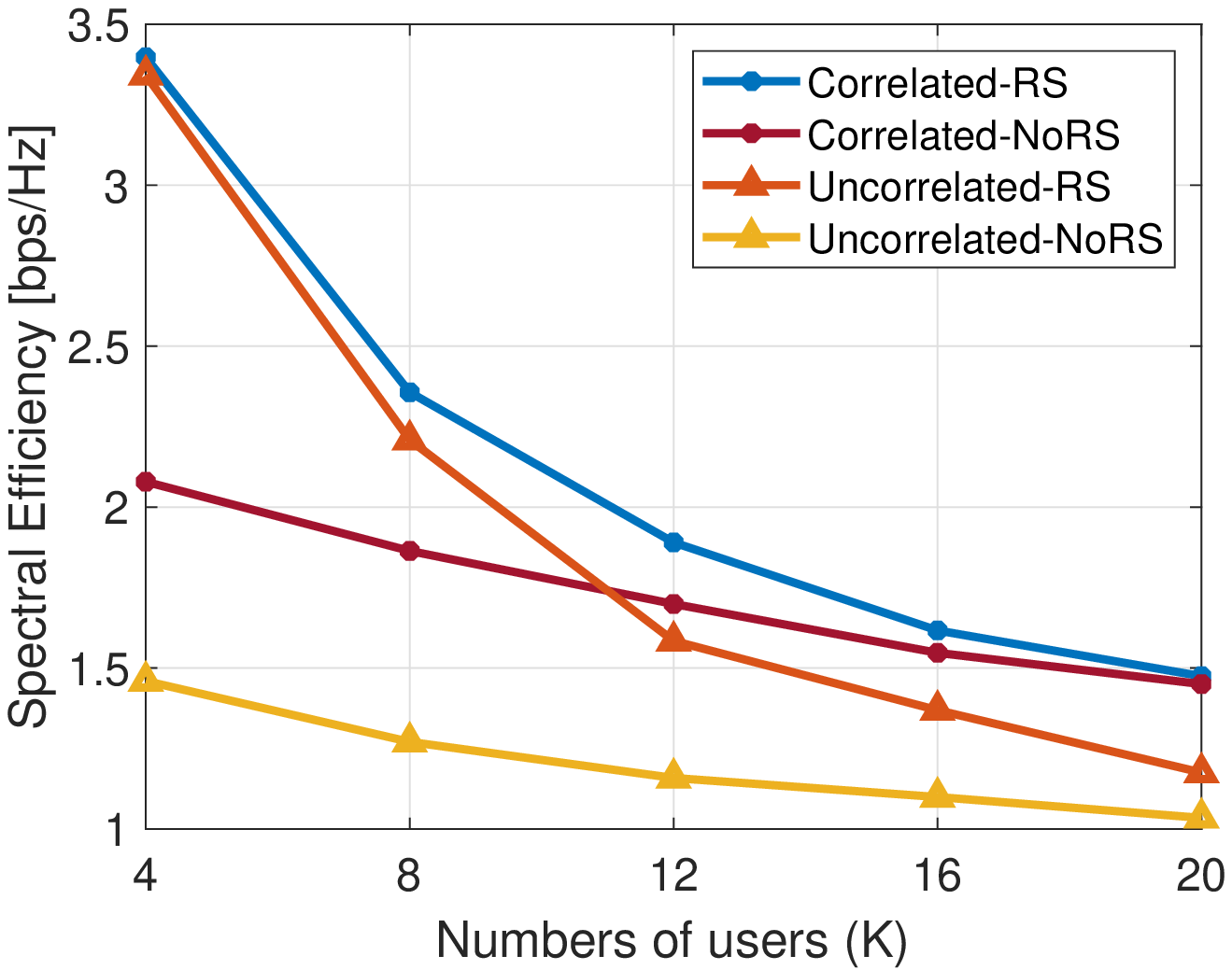}}
\subfloat[MaxSINR]{\includegraphics[width=4.6cm,height=4.7cm]{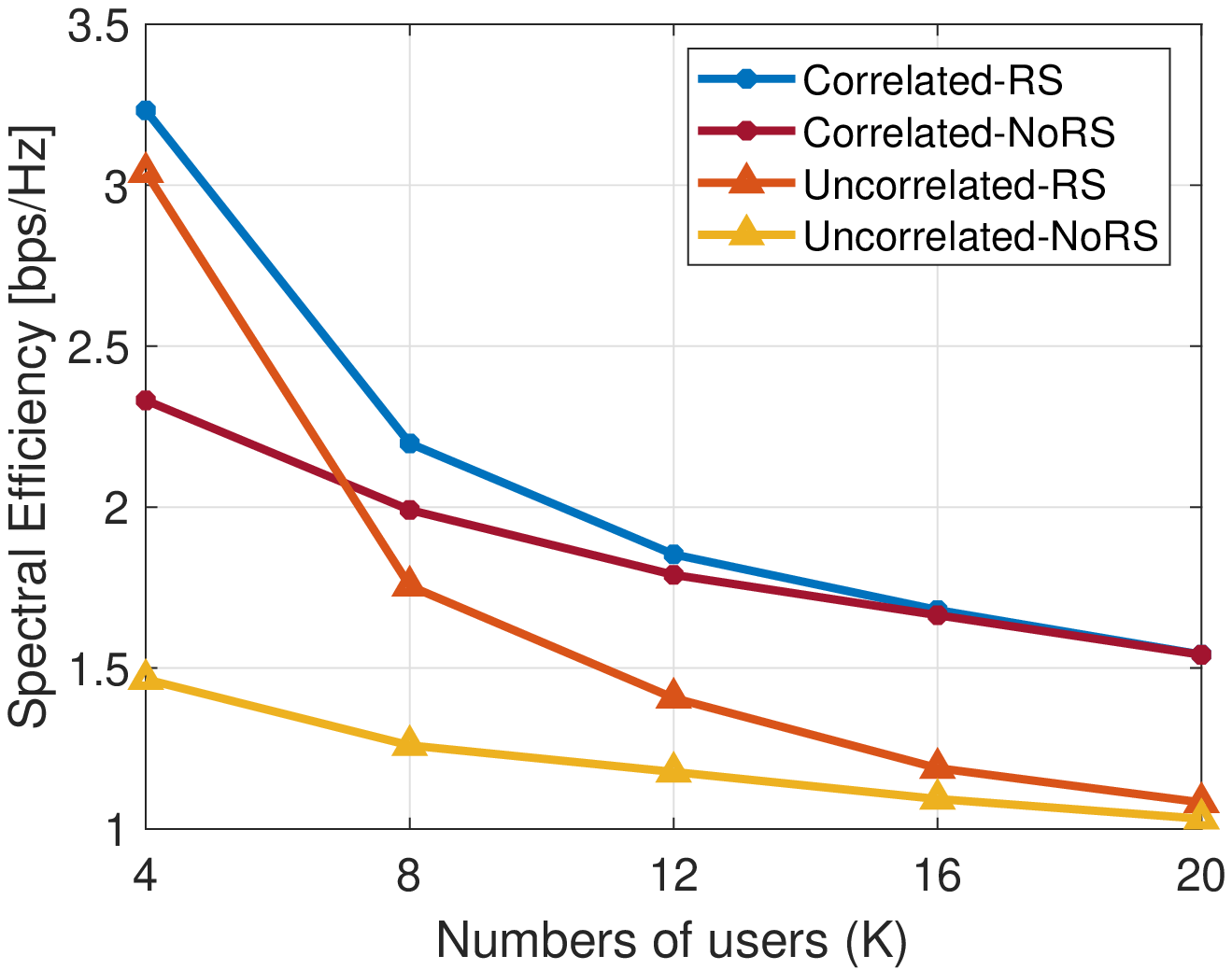}}
\caption{ Sum-SE performance of the two low-complexity power allocation schemes in circular topology, $M=100$.}
\label{fig:MaxSumSE_MaxSINR}
\end{figure}

In this subsection, we illustrate the sum-SE performance of both RS and NoRS strategies with low-complexity power allocation schemes, which are more beneficial for use cases like mMTC and URLLC. {Similar to the MaxMin power allocation scheme, RS and NoRS transmission strategies achieve the same SE performance in the absence of pilot contamination, with both MaxSum-SE and MaxSINR. Therefore, for brevity, we do not illustrate the SE performance with no pilot contamination.} Moreover, taking note of the inferences from MaxMin results, we begin by focusing on the network layout with significant gains for RS, i.e., circular topology with UEs confined within a sector of $\theta=\pi/4$. From Fig.~\ref{fig:MaxSumSE_MaxSINR}(a) and Fig.~\ref{fig:MaxSumSE_MaxSINR}(b), we observe that the SE performance of RS and NoRS for both power allocation schemes is consistent with the MaxMin power allocation scheme. While RS always achieves a higher sum-SE than NoRS, the gap is lower with spatially correlated channels. For $K=8$, the SE gain of RS over NoRS with MaxSum-SE when UE channels are spatially correlated and uncorrelated is $20.9\%$ and $42\%$, respectively. Both RS and NoRS achieve a lower sum-SE with MaxSINR compared to MaxSum-SE as the number of UEs increases. The lower sum-SE performance is because implicitly MaxSINR is designed to strike a balance between maximizing sum-SE and fairness. As a result, MaxSINR always allocates a non-zero power to each UE, resulting in sum-SE performance loss. As observed with the MaxMin scheme, the gain of RS over NoRS will decrease when the UEs are not confined to a sector and instead are randomly distributed over the entire circle. Similarly, the gap between the sum-SE of RS and NoRS will decrease in rectangular topology, i.e., when UEs experience different path loss. Nonetheless, despite low-complexity power allocation schemes, RS is robust to imperfect CSIT and achieves a better SE performance than NoRS.
\subsection{{Impact of number of transmit antennas}}
\begin{figure}
\centering
\subfloat[]{\includegraphics[width=4.6cm,height=4.7cm]{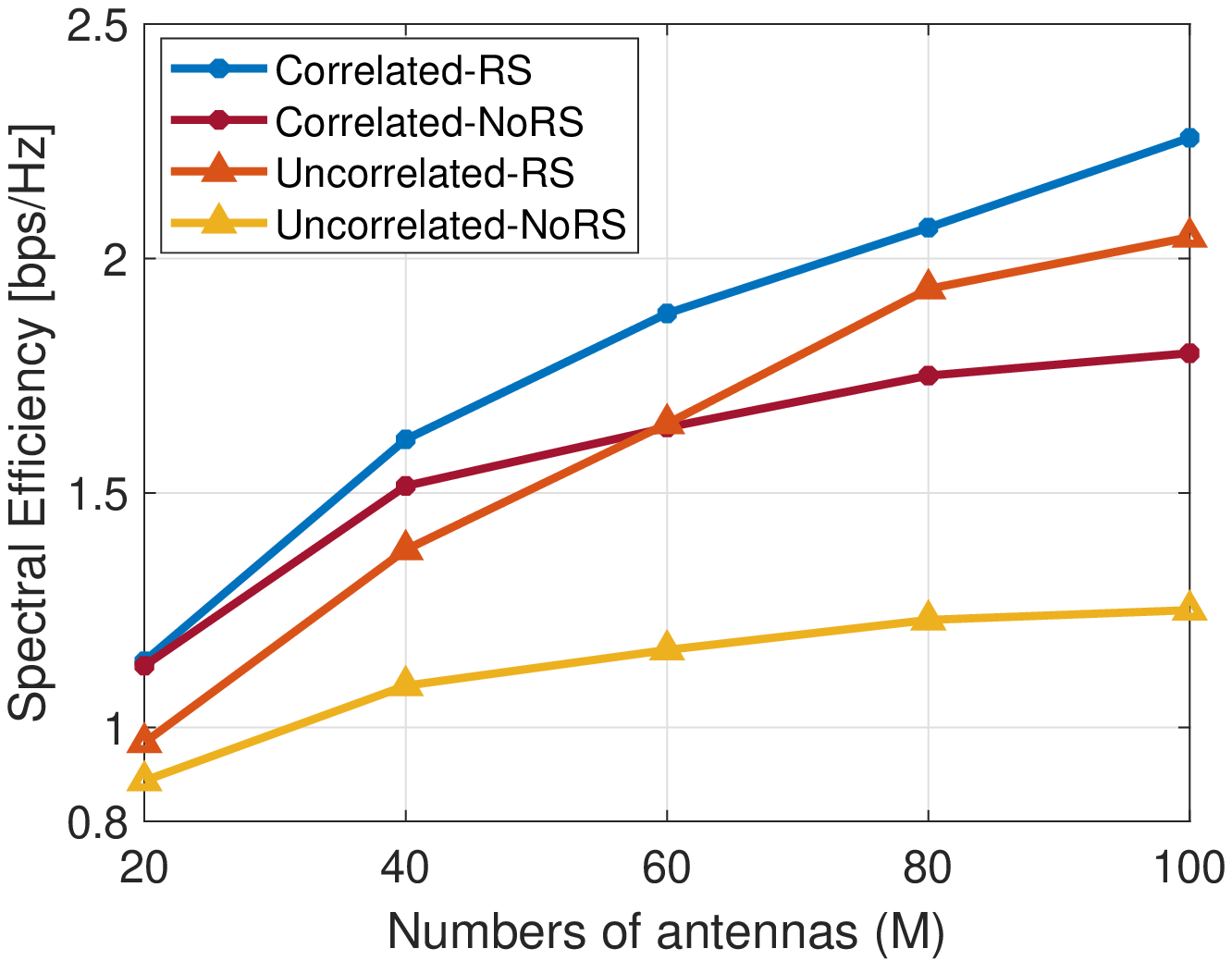}}
\subfloat[]{\includegraphics[width=4.6cm,height=4.7cm]{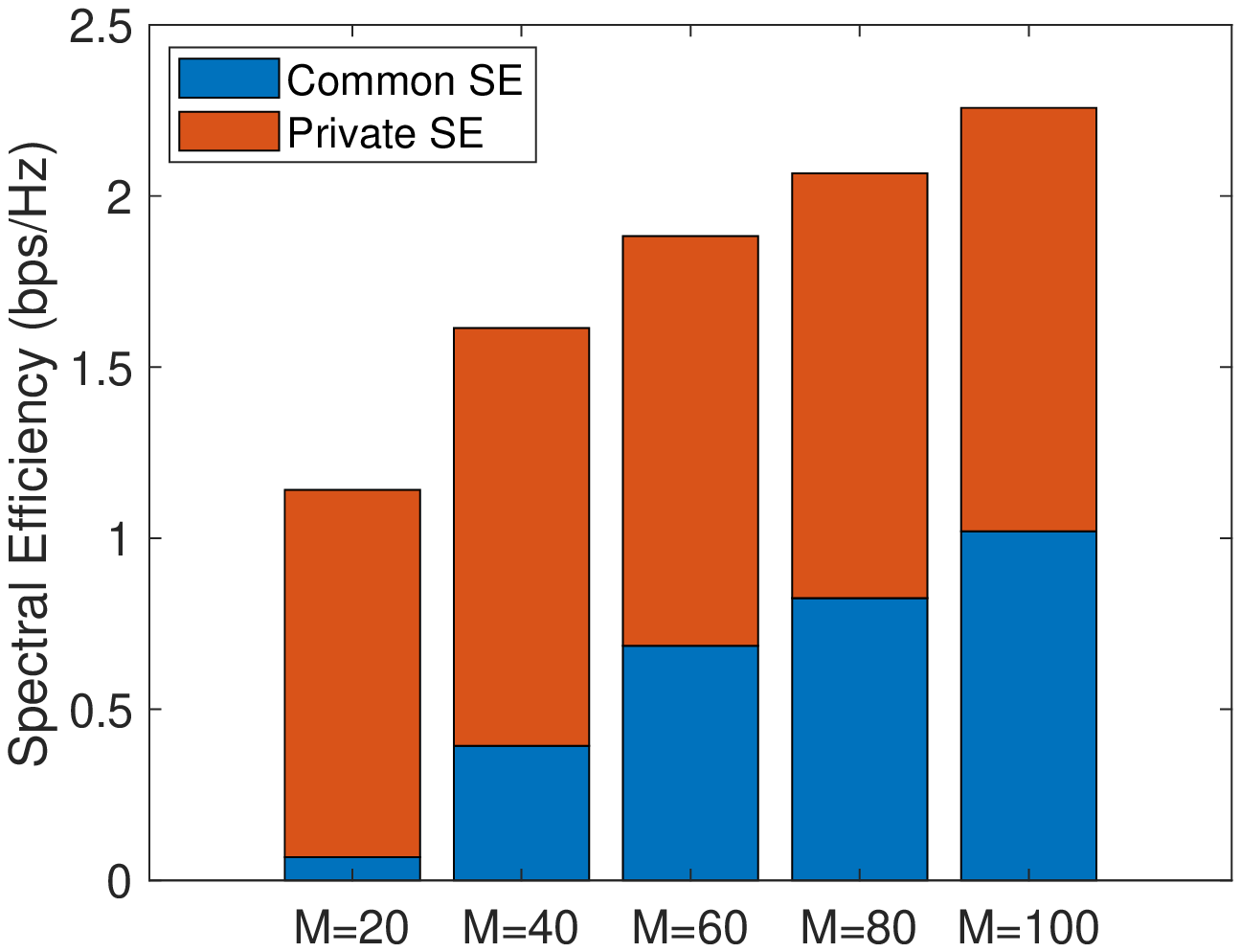}}
\caption{ Sum-SE performance versus $M$, $K=8$.}
\label{fig:TransvsAnt}
\end{figure}

For analysis, we continue with MaxSum-SE power allocation scheme and keep the network layout the same as Fig.~\ref{fig:Topology_Diag_Pi4}(b).  Fig.~\ref{fig:TransvsAnt}(a) illustrates the sum-SE as a function of the number of transmit antennas, $M$, with the number of UEs $K=8$. We observe that RS mitigates the effect of pilot contamination even with a finite number of antennas, and as $M$ increases, the sum-SE gain of RS over NoRS increases. Since the effective propagation channel between the BS and UE $k$ provides asymptotic channel hardening as $M\rightarrow\infty$, it aids in the SIC of the common stream. Therefore, as channel hardening increases, the performance gain of RS over NoRS increases due to better management of interference at the UE side. Note that the increase in the performance gain is more explicit with spatially uncorrelated channels than correlated channels. This behaviour can be attributed to the fact that spatial correlation undermines channel hardening in MaMIMO \cite{Saptial@hardening},  thereby disallowing both RS and NoRS to exploit the full potential of MaMIMO. Nonetheless, we observe from Fig.~\ref{fig:TransvsAnt}(b) that the contribution of common SE to the total sum-SE with spatially correlated channels increases from approximately $6\%$ to $45\%$ when $M$ increases from $20$ to $100$. Therefore, the SE gain of RS over NoRS increases with $M$.
\section{Conclusion}\label{Concl}
In this paper, we proposed a general DL transmission framework of RSMA in TDD MaMIMO network. Based on the proposed framework, lower SE bounds for the common and private streams were derived that hold true for any choice of channel estimation and precoder design. Moreover, a low-complexity precoder design was formulated for the common stream of RS. We devised power allocation schemes of RS for three different network utility functions such that the formulation holds for any choice of UL channel estimator and DL precoders. Considering both the ideal case when UEs use orthogonal pilots, and the worst case when multiple UEs share the same pilot, we analyzed the SE performance of RS and NoRS. Through numerical simulations, we illustrated the SE performance of RS and compared it with that of a NoRS strategy in different network topologies and propagation environments. Numerical results showed that RSMA is significantly more robust to pilot contamination than a conventional SDMA-based MaMIMO transmission strategy. 
\par The scarcity of pilots will be a major concern in B$5$G and $6$G networks, given their peculiar characteristics and use cases. Because of the adaptability and robustness of RSMA under different CSIT regimes, RSMA will potentially play a significant role in MaMIMO networks. Building on the work done in this paper, future works can focus on studying multigroup multicasting in MaMIMO with RSMA and investigating RSMA as an inter-cell pilot contamination mitigation strategy. Furthermore, a majority of the literature (including this paper) studying performance analysis of conventional TDD MaMIMO network predominantly assumes the availability of channel statistics at the BS. The assumption allows for employing sophisticated signal processing algorithms like MMSE estimation to mitigate the effect of pilot contamination. However, in the absence of the knowledge of channel statistics, the difficulty of dealing with pilot contamination would increase significantly. Therefore, it would be interesting to study the potential benefits of RSMA under such constraints. Future works of RSMA in TDD MaMIMO could also focus on the interplay of RSMA with different UE grouping algorithms, UL performance of RSMA, and performance of RSMA in TDD MaMIMO with DL training. {Finally, beyond the 1-layer RS architecture utilized here, other RS schemes can be leveraged to further enhance the SE performance.}
\ifCLASSOPTIONcaptionsoff
  \newpage
\fi

\appendices
\section{}\label{App_A}
 We obtain the closed-form expression of the entity $\mathbb{E}\{\mathbf{g}_{k}^{H}\mathbf{w}_{c}^{*}\}$ as 
\begin{equation}\label{eq:ClosedForm_Wc}
\mathbb{E}\{\mathbf{g}_{k}^{H}\mathbf{w}_{c}^{*}\}=\frac{\sum_{i=1}^{K}a_{i}^{*}tr(\mathbf{R}_{i}\mathbf{Q}^{-1}\mathbf{R}_{k})}{\sqrt{\sum_{i=1}^{K}\sum_{j=1}^{K}a_{i}^{*}a_{j}^{*}tr(\mathbf{R}_{i}\mathbf{Q}^{-1}\mathbf{R}_{j})}}.
\end{equation}
Similarly, the closed-form expression of $\mathbb{E}\{|\mathbf{g}_{k}^{H}\mathbf{w}_{c}^{*}|^{2}\}$ is obtained as
\begin{equation}\label{eq:Closedform_gkwc}
\mathbb{E}\{|\mathbf{g}_{k}^{H}\mathbf{w}_{c}^{*}|^{2}\}=\frac{\sum_{i=1}^{K}\sum_{j=1}^{K}a_{i}^{*}a_{j}^{*}\mathbb{E}\{\mathbf{g}_{k}^{H}\widehat{\mathbf{g}}_{i}\widehat{\mathbf{g}}_{j}^{H}\mathbf{g}_{k}\}}{\sum_{i=1}^{K}\sum_{j=1}^{K}a_{i}^{*}a_{j}^{*}tr(\mathbf{R}_{i}\mathbf{Q}^{-1}\mathbf{R}_{j})}.
\end{equation}
Using (\ref{eq:Dependent_Ch}), numerator in equation (\ref{eq:Closedform_gkwc}) is reduced to
\begin{equation}\label{eq:ReducedClosedform_wc}
\begin{split}
\mathbb{E}\{\mathbf{g}_{k}^{H}\widehat{\mathbf{g}}_{i}\widehat{\mathbf{g}}_{j}^{H}\mathbf{g}_{k}\}=&\,\mathbb{E}\{\mathbf{g}_{k}^{H}\widehat{\mathbf{g}}_{i}\widehat{\mathbf{g}}_{i}^{H}\mathbf{R}_{i}^{-1}\mathbf{R}_{j}\mathbf{g}_{k}\}\\
=&\,tr(\mathbf{R}_{i}^{-1}\mathbf{R}_{j}\mathbb{E}\{\mathbf{g}_{k}^{H}\widehat{\mathbf{g}}_{i}\widehat{\mathbf{g}}_{i}^{H}\mathbf{g}_{k}\}).
\end{split}
\end{equation}
As $\mathbf{g}_{k}=\widehat{\mathbf{g}}_{k}+\widetilde{\mathbf{g}}_{k}$, and $\widehat{\mathbf{g}}_{k}$ and estimation error $\widetilde{\mathbf{g}}_{k}$ are independent, (\ref{eq:ReducedClosedform_wc}) is further reduced to
\begin{equation}\label{eq:ReducedClosedform_wc2}
\begin{split}
\mathbb{E}\{\mathbf{g}_{k}^{H}\widehat{\mathbf{g}}_{i}\widehat{\mathbf{g}}_{j}^{H}\mathbf{g}_{k}\}=&\,tr\big(\mathbf{R}_{i}^{-1}\mathbf{R}_{j}\mathbb{E}\{\widehat{\mathbf{g}}_{k}^{H}\widehat{\mathbf{g}}_{i}\widehat{\mathbf{g}}_{i}^{H}\widehat{\mathbf{g}}_{k}\}\big)\\
&+tr\big(\mathbf{R}_{i}^{-1}\mathbf{R}_{j}\mathbb{E}\{\widetilde{\mathbf{g}}_{k}\widetilde{\mathbf{g}}_{k}^{H}\}\mathbb{E}\{\widehat{\mathbf{g}}_{i}\widehat{\mathbf{g}}_{i}^{H}\}\big).
\end{split}
\end{equation}
Using equation (C.64) and equation (C.65) in \cite{massivemimobook}, we get
\begin{equation}\label{eq:ReducedClosedform_wc3}
\begin{split}
&\mathbb{E}\{\widehat{\mathbf{g}}_{k}^{H}\widehat{\mathbf{g}}_{i}\widehat{\mathbf{g}}_{i}^{H}\widehat{\mathbf{g}}_{k}\}= tr(\boldsymbol{\Phi}_{i}\boldsymbol{\Phi}_{k})+|tr(\mathbf{R}_{i}\mathbf{Q}^{-1}\mathbf{R}_{k})|^2,\\
&\mathbb{E}\{\widetilde{\mathbf{g}}_{k}\widetilde{\mathbf{g}}_{k}^{H}\}\mathbb{E}\{\widehat{\mathbf{g}}_{i}\widehat{\mathbf{g}}_{i}^{H}\}=(\mathbf{R}_{k}-\boldsymbol{\Phi}_{k})\boldsymbol{\Phi}_{i},
\end{split}
\end{equation}
where $\mathbf{R}_{k}$ and $\boldsymbol{\Phi}_{k}$ are the correlation matrices of the channel and channel estimate of UE $k$, respectively. Substituting equation (\ref{eq:ReducedClosedform_wc3}) into  (\ref{eq:ReducedClosedform_wc2}) and, in turn, substituting (\ref{eq:ReducedClosedform_wc2}) into (\ref{eq:Closedform_gkwc}), closed-form expression for $\mathbb{E}\{|\mathbf{g}_{k}^{H}\mathbf{w}_{c}^{*}|^{2}\}$ is obtained.
\section{}\label{App_B}
\subsection{MaxSum-SE: SCA}\label{MaxSumSE_SCA_alg}
Similar to the problem (\ref{eq:MaxMin_2}), we introduce auxiliary variable $\alpha_{c}$ and vector $\boldsymbol{\alpha_{p}}=[\alpha_{p,1},\ldots,\alpha_{p,K}]$ representing the common SE and private SEs of UEs, respectively. We introduce $\mathbf{r}_{c}=[r_{c,1},\ldots,r_{c,K}]$ and $\mathbf{r}_{p}=[r_{p,1},\ldots,r_{p,K}]$, representing $1$ plus SINR value for common and private streams of UEs respectively and transform problem (\ref{eq:MaxSumSE}) equivalently as
\begin{subequations}\label{eq:MaxSumSE_1}
\begin{align}
\max_{\substack{\boldsymbol{\rho},\boldsymbol{\alpha}_{p},{\alpha}_{c},\\\mathbf{r}_{p},\mathbf{r}_{c}}}&\;\;\;\alpha_{c}+\sum_{k=1}^{K}\alpha_{p,k},\\ 
s.t.\;\;\;& r_{p,k} - 2^{\frac{\tau}{\tau_{d}}\alpha_{p,k}} \geq 0 \;\;\; \forall k\in\mathcal{K},\\ 
& r_{c,k} - 2^{\frac{\tau}{\tau_{d}}\alpha_{c}} \geq 0 \;\;\; \forall k\in\mathcal{K},\\ 
& \frac{\rho_c a_{c,
k}}{\sum_{i=1}^{K}\rho_i b_{ki}^{c}+\rho_{c}I_{c,k}+\sigma_{n}^{2}} \geq r_{c,k}-1,\,\forall k\in\mathcal{K},\\ 
& \frac{\rho_k a_{p,k}}{\sum_{i=1}^{K}\rho_i b_{ki}^{p}+\rho_{c}I_{c,k}+\sigma_{n}^{2}} \geq r_{p,k}-1,\,\forall k\in\mathcal{K},\\ 
&\;\;\;\;\;\alpha_{c} \geq 0,\\ 
& \rho_{c} + \sum_{i=1}^{K} \rho_{i} \leq \rho_{\textrm{dL}}.
\end{align}
\end{subequations}
We can follow the approach used to solve the MaxMin problem (\ref{eq:MaxMin_2}) and obtain the power allocation coefficients that aims to maximize the sum-SE. {To avoid redundancy, we do repeat the procedure of obtaining the SCA algorithm for MaxSum-SE. Let us denote the SCA-based MaxSum-SE power allocation algorithm of RS as ``MaxSumSE-SCA''. With $2(3K+1)$ variables and constraints, the computational complexity of MaxSumSE-SCA will be of the order of
\begin{equation}\label{eq:maxsumSE_complexity}
    \mathcal{O}\left(N_{1}\max\left\{8\left(3K+1\right)^{3},\,F_{3}\right\}\right),
\end{equation}
where $F_{3}$ is the cost of evaluating the first and second
derivatives of the objective and constraint functions of the MaxSumSE-SCA algorithm. $N_{1}$ is the number of iterations required for the algorithm to converge. From (\ref{eq:maxsumSE_complexity}), we observe that the complexity of MaxSumSE-SCA is significantly higher compared to Algorithm \ref{alg:MaxSumSE_alg}, even for small values of $K$ and $\Delta$. For ease of illustration and understanding, Table \ref{tab:Comp_table_MaxSumSE} compares the MaxSumSE-SCA method and Algorithm \ref{alg:MaxSumSE_alg} in terms of the CPU time\footnote{{The simulations were performed on a Windows $10$ workstation with $3.00$ GHz i$7-9700$ CPU and $8$ GB of random access memory.}} consumed to obtain the sum-SE performance with a single pilot used by $K=4$ UEs for UL training. We observe that the MaxSumSE-SCA  scheme achieves $30.7\%$ higher sum-SE performance but the time taken increases approximately $10^{5}$ fold. Note that the time consumed and sum-SE performance of MaxSumSE-SCA also depends on the power allocation initialization. For comparison, in Table \ref{tab:Comp_table_MaxSumSE}, we initialize $\boldsymbol{\rho}^{[0]}$ as $\rho_{c}=0.1\rho_{\textrm{dL}}$ and $\rho_{k}={0.9\rho_{\textrm{dL}}}/{K},\, \forall k\in\mathcal{K}$.   
\begin{table}[!ht]
\centering
	\caption{MaxSum-SE: Performance Comparison}
	    \label{tab:Comp_table_MaxSumSE}
	\begin{tabular}{|L{3cm} | C{2cm} | C{2cm}|}
		\hline
		 & \textbf{Algorithm \ref{alg:MaxSumSE_alg}} & \textbf{SCA} \\
		\hline
   CPU time (secs)& $0.0013$ & $86.94$\\
        \hline
   SE (bps/Hz) & $3.47$ & $5.01$ \\
		\hline
	\end{tabular}
\end{table}}
\vspace{-0.5cm}
\subsection{MaxSINR: SCA}\label{MaxSINR_SCA_alg}
We introduce auxiliary variable ${\alpha_{c}}$ and auxiliary vector $\boldsymbol{\alpha_{p}}=[\alpha_{p,1},\ldots,\alpha_{p,K}]$ representing the log of common SINR and private SINRs of UEs, respectively. Similarly, we introduce $\mathbf{r}_{c}=[r_{c,1},\ldots,r_{c,K}]$ and $\mathbf{r}_{p}=[r_{p,1},\ldots,r_{p,K}]$, representing SINR value for common and private streams of UEs respectively and transform problem (\ref{eq:MaxSINR_1}) equivalently as
\begin{equation}\label{eq:MaxSINR_App}
\begin{split}
\max_{\substack{\boldsymbol{\rho},\boldsymbol{\alpha}_{p},{\alpha}_{c},\\\mathbf{r}_{p},\mathbf{r}_{c}}}&\;\;\;\alpha_c+\sum_{k=1}^{K}\alpha_{p,k},\\ 
s.t.\;\;\;\;\;\;&\frac{\rho_c a_{c,
k}}{\sum_{i=1}^{K}\rho_i b_{ki}^{c}+\rho_{c}I_{c,k}+\sigma_{n}^{2}} \geq r_{c,k},\,\forall k\in\mathcal{K},\\ 
& \frac{\rho_k a_{p,k}}{\sum_{i=1}^{K}\rho_i b_{ki}^{p}+\rho_{c}I_{c,k}+\sigma_{n}^{2}} \geq r_{p,k},\,\forall k\in\mathcal{K},\\ 
&(\ref{eq:MaxSumSE_1}b),\,(\ref{eq:MaxSumSE_1}c),\,(\ref{eq:MaxSumSE_1}f),\,(\ref{eq:MaxSumSE_1}g).
\end{split}
\end{equation}
{Let us denote the SCA-based MaxSINR power allocation algorithm of RS as ``MaxSINR-SCA''. The complexity of  MaxSINR-SCA will be the same as of MaxSumSE-SCA and is delineated in equation (\ref{eq:maxsumSE_complexity}). Observing (\ref{eq:maxSINR_complexity}) and (\ref{eq:maxsumSE_complexity}), the complexity of MaxSINR-SCA is higher than that of Algorithm \ref{alg:MaxSINR_alg} and as $K$ increases, the gap in complexity grows significantly}. From Table \ref{tab:Comp_table_MaxSINR}, for $K=4$, we observe that MaxSINR-SCA achieves approximately $20\%$ lower sum-SE performance and consumes $8$ fold more time compared to Algorithm \ref{alg:MaxSINR_alg}. {Low complexity, specific form of the objective and constraints, and modern solvers allows Algorithm \ref{alg:MaxSINR_alg} based on GP to efficiently optimize the power coefficients and find an optimal solution in significantly less time\cite{hoburg2014geometric}}. As a result, Algorithm \ref{alg:MaxSINR_alg} outperforms MaxSINR-SCA in terms of both SE and time consumed to compute the sum-SE. For MaxSINR-SCA as well, both the time and sum-SE performance depend on the initialization of the power coefficients, which are taken as in Appendix \ref{MaxSumSE_SCA_alg}. 
\begin{table}[!ht]
\centering
	\caption{MaxSINR: Performance Comparison}
	    \label{tab:Comp_table_MaxSINR}
	\begin{tabular}{|L{3cm} | C{2cm} | C{2cm}|}
		\hline
		 & \textbf{Algorithm \ref{alg:MaxSINR_alg}} & \textbf{SCA} \\
		\hline
   CPU time (secs)& $4.33$ & $34.10$\\
        \hline
   SE (bps/Hz) & $3.24$ & $2.62$ \\
		\hline
	\end{tabular}
\end{table}

\bibliographystyle{IEEEtran}
\bibliography{reference}
\end{document}